\documentclass[USenglish,oneside,twocolumn]{article}

\usepackage[utf8]{inputenc}
\usepackage[big]{dgruyter_NEW}

\usepackage{tikz}
\usepackage{graphicx}
\usepackage{booktabs}
\usepackage{color}
\usepackage{xcolor}
\usepackage{varwidth}
\usepackage{colortbl}
\usepackage{longtable}
\usepackage{verbatim}
\usepackage{forloop,supertabular}
\usepackage[flushleft]{threeparttable}
\usepackage{tabularx}
\usepackage{placeins}
\usepackage{nowidow}
\usepackage{float}
\usepackage{tabularx}

\usetikzlibrary{scopes}
\usetikzlibrary{positioning}
\usetikzlibrary{shapes}
\usepackage[misc,clock,geometry]{ifsym}
\usepackage{placeins}

\newcommand{\PDIa}{It was incomprehensible to select cookie settings.}
\newcommand{\PDIb}{It was frustrating to select cookie settings.}
\newcommand{\PDIc}{It was easy to select cookie settings.}
\newcommand{\PDEa}{When it comes to cookie settings, the website is dishonest towards its users.}
\newcommand{\PDEb}{The website tries to mislead users towards selecting cookie settings which they do not intend to select.}
\newcommand{\PDEc}{The website makes use of misleading tactics so that users select cookie settings which they do not intend to select.}
\newcommand{\RGa}{I regret my choice of cookie settings.}
\newcommand{\RGb}{I would change my cookie settings if it was possible.}
\newcommand{\RGc}{I am satisfied with my choice of cookie settings.}
\newcommand{\PAa}{It is important for me to protect my privacy online.}
\newcommand{\PAb}{If websites use cookies, my online privacy is impaired.}
\newcommand{\PAc}{I am concerned about my online privacy being impaired by website cookies.}

\newcommand{\GERPDIa}{Es war unverst\"andlich, eine Auswahl zu treffen.}
\newcommand{\GERPDIb}{Es war frustrierend, eine Auswahl zu treffen.}
\newcommand{\GERPDIc}{Es war einfach, eine Auswahl zu treffen.}
\newcommand{\GERPDEa}{Die Seite ist bez\"uglich der Cookie-Einstellungen unehrlich gegen\"uber ihren Nutzern.}
\newcommand{\GERPDEb}{Die Seite versucht Nutzer dazu zu f\"uhren, Cookie-Einstellungen zu w\"ahlen, die sie nicht w\"ahlen wollen.}
\newcommand{\GERPDEc}{Die Seite benutzt irref\"uhrende Taktiken, damit Nutzer Cookie-Einstellungen w\"ahlen, die sie nicht w\"ahlen wollen.}
\newcommand{\GERRGa}{Ich bereue meine getroffene Auswahl.}
\newcommand{\GERRGb}{Ich w\"urde meine Auswahl \"andern, wenn ich die M\"oglichkeit h\"atte.}
\newcommand{\GERRGc}{Ich bin mit meiner Auswahl zufrieden.}  
\newcommand{\GERPAa}{Der Schutz meiner Privatsph\"are im Internet ist mir wichtig.}
\newcommand{\GERPAb}{Wenn Webseiten Cookies verwenden, schr\"ankt dies meine Privatsph\"are ein.}
\newcommand{\GERPAc}{Ich bin besorgt dar\"uber, dass meine Privatsph\"are durch Cookies von Webseiten eingeschr\"ankt wird.}
 
\DOI{foobar}

\cclogo{\includegraphics{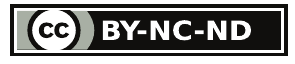}}

\begin{document}

  \author*[1]{Dominique Machuletz}

  \author[2]{Rainer Böhme}

  \affil[1]{Independet, E-mail: mail@machuletz.com. Work carried out while at the University of Münster, Germany.}

  \affil[2]{University of Innsbruck, Austria, E-mail: rainer.boehme@uibk.ac.at}

  \title{\huge Multiple Purposes, Multiple Problems:\\A User Study of Consent Dialogs after GDPR}

  \runningtitle{A User Study of Consent Dialogs after GDPR}


  \begin{abstract}
{
The European Union's General Data Protection Regulation (GDPR) requires websites to ask for consent to the use of cookies for \emph{specific purposes}. 
This enlarges the relevant design space for consent dialogs. 
Websites could try to maximize click-through rates and positive consent decision, even at the risk of users agreeing to more purposes than intended. 
We evaluate a practice observed on popular websites by conducting an experiment with one control and two treatment groups ($N=150$ university students in two countries). 
We hypothesize that users' consent decision is influenced by (1) the number of options, connecting to the theory of choice proliferation, and (2) the presence of a highlighted default button (``select all''), connecting to theories of social norms and deception in consumer research. 
The results show that participants who see a default button accept cookies for more purposes than the control group, while being less able to correctly recall their choice.  
After being reminded of their choice, they regret it more often and perceive the consent dialog as more deceptive than the control group. 
Whether users are presented one or three purposes has no significant effect on their decisions and perceptions. We discuss the results and outline policy implications.
}
\end{abstract}

  \keywords{web privacy, user study, consent, cookies, controlled experiment, choice proliferation, deception, privacy paradox, privacy by design, dark patterns}

  \journalname{Proceedings on Privacy Enhancing Technologies}
\DOI{10.2478/popets-2020-0037}
  \startpage{481}
  \received{2019-08-31}
  \revised{2019-11-04} 
  \accepted{2019-11-25}

  \journalyear{}
  \journalvolume{2020}
  \journalissue{2}

\maketitle

\section{Introduction}

The European Union's General Data Protection Regulation (GDPR)~\cite{gdpr} came into force in May 2018.
It stipulates that \emph{data controllers} (e.\,g., website operators) must have a legal basis for the collection and processing of  \emph{personal data}. One legal basis is \emph{consent}: \emph{data subjects} (users) agree to the data processing for \emph{specific purposes}. 
While these requirements are not new,\footnote{The principles of consent and purpose binding appear in data protection laws since the 1970s. The specific case for web cookies was harmonized in the EU through the 2009 update of the ePrivacy Directive~\cite{epd,epdupdate,leenes2015}, but respected by only one in two websites, according to a recent measurement study~\cite{trevisan2019pets}.} the GDPR's threat of sanctions and more effective enforcement led many website operators to rethink their cookie practices, or at least ensure compliance by obtaining consent before using cookies for purposes that are not covered by other legal bases~\cite{vanEijk2019}. 

Web cookies are key--value pairs stored on the client device for purposes ranging from session tracking, user recognition, counting unique users, third-party tracking to profiling and targeted advertising~\cite{englehardt2016}. As every cookie can in principle serve many purposes at the same time, and \emph{necessary} cookies not carrying any personal data do not require consent, a user generally cannot verify if a website complies with the agreed purposes.

Common methods for asking web users to decide on the cookie settings are pop-up banners or dialogs that appear at the beginning of each user's first visit of a website. They typically include a notice on the data collection that asks users whether they consent to (parts of) the practices. Systematic longitudinal measurements are lacking, but one study reports that 62\% of the websites in its sample used such notices in June 2018~\cite{degeling2018we}. It also shows that the implementation---specifically, the granularity of control offered to users---differs between websites. The authors of~\cite{degeling2018we} conjecture that many cookie banners and dialogs are not very usable, and they provide early evidence from a series of field experiments with several variants of cookie banners placed on one website~\cite{utz2019ccs}.

\begin{figure}
	\begin{center}
		\includegraphics[width=\linewidth]{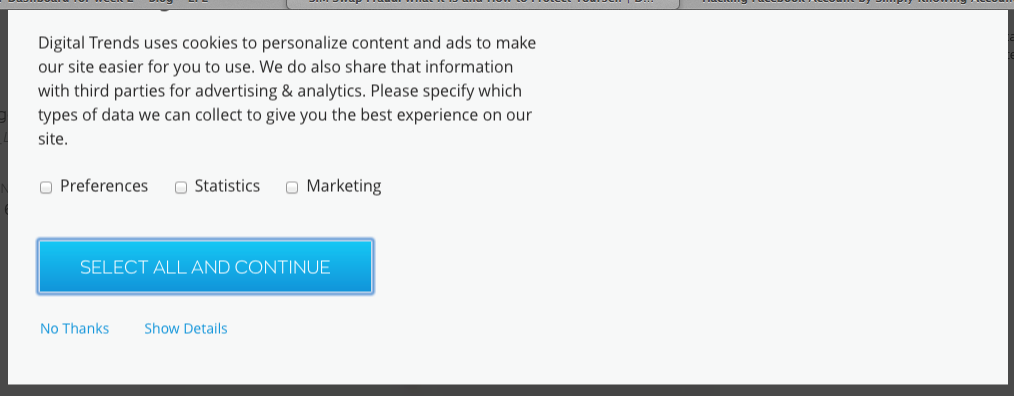}
		\includegraphics[width=\linewidth]{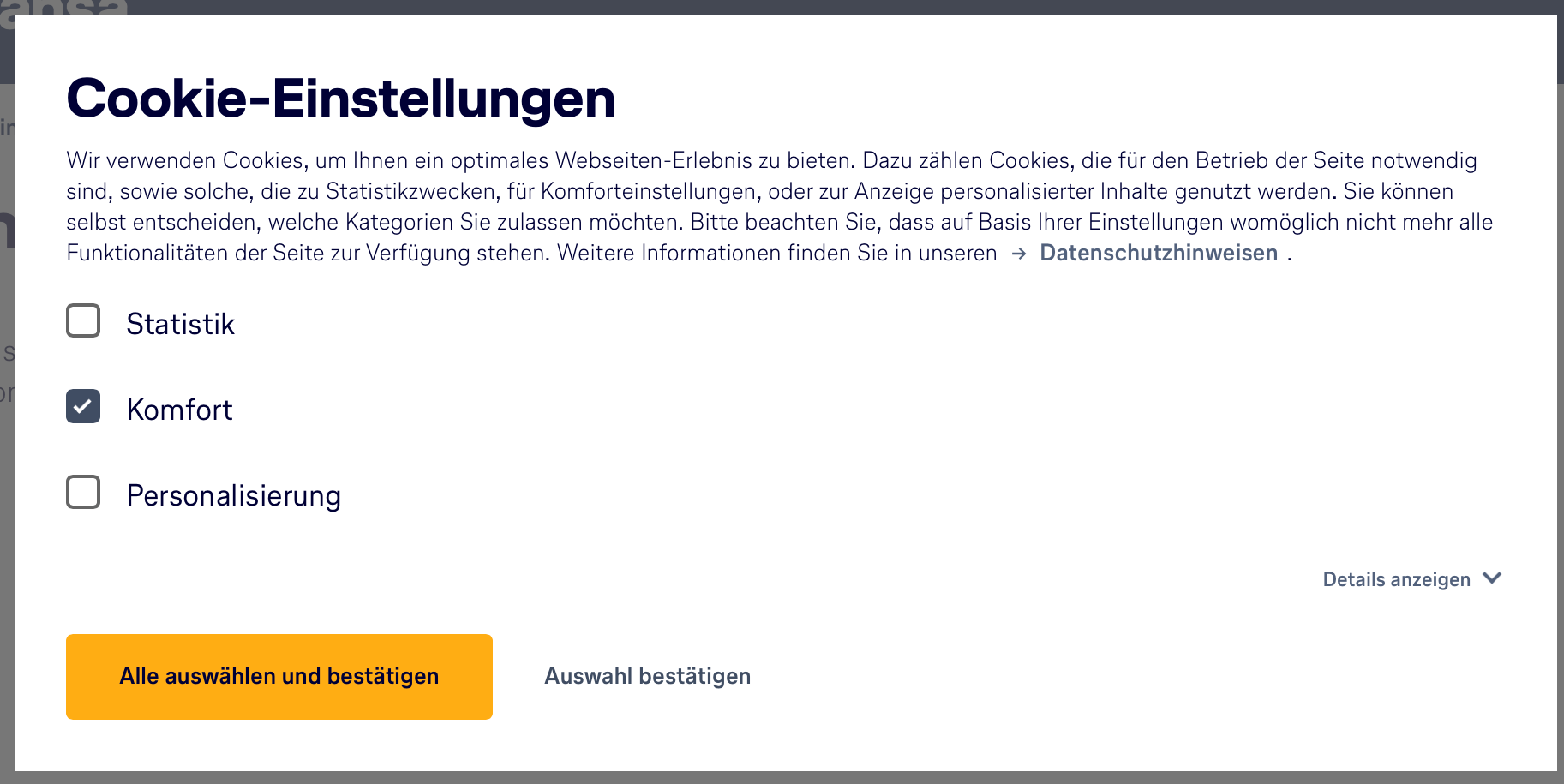}
	\end{center}
	\caption{\label{fig:banner} Examples of real-world cookie consent dialogs that motivated this study: a US technology news website (top) and a German airline website (bottom). Both dialogs are blocking and all items are unchecked initially (opt-in).}
\end{figure}

Independently, in November 2018, we noticed that some cookie consent dialogs seem to be designed to nudge users into accepting all displayed purposes. (This observation is meanwhile documented in the literature~\cite[e.\,g.,][]{Sanchez-Rola2019}). It is understandable that the industry finds cookie banners disadvantageous as they add friction to the user experience and might limit the ability to track users on and across sites. Hence, there is ample business interest in minimizing friction and maximizing positive consent decisions by optimizing interface design. Common design elements in the dialogs we observed (see Figure~\ref{fig:banner} for examples) are checkboxes for several purposes of data processing as well as buttons to either select all purposes at once or to confirm the manual selection before accessing the website.

We identify two features that might compromise usability. First, the highlighted button automatically accepts all purposes, regardless of whether any checkboxes have (or have not) been selected before the button is clicked. This button does not increase the users' choice options, but might rather ``trick'' them into accepting all purposes without actively selecting them. Second, the number of selectable purposes may influence users' choice as former studies in the field of psychology revealed that a high number of alternatives has adverse effects on individuals' decision making~\cite{kling2008misperception,cronqvist2004design}. This phenomenon has also been demonstrated in the context of privacy settings~\cite{korff2014too}. 

These considerations call for a user study, which we have carried out in the form of a controlled classroom experiment and report in this paper. Our general research question is:
\\\\
\noindent \textit{How do users react to design features of multi-purpose consent dialogs on the web in terms of actual behavior and stated perceptions?}
\\\\
\indent The rest of this paper is structured as follows. First, we review the literature on consent dialog designs in Section~\ref{sec:relatedwork}. Section~\ref{sec:theory} recalls the theoretical background on choice proliferation and deception, from which we derive our hypotheses. The instrument and the administration of the controlled experiment is described in Section~\ref{sec:method}. The results of our hypothesis tests (Section~\ref{sec:results}) precede the discussion of our findings (Section~\ref{sec:discussion}). We conclude with some recommendations for interface design 
and policy development
in Section~\ref{sec:conclusion}.

\section{Background}
\label{sec:relatedwork}


We first summarize the legal requirements for GDPR-compliant consent dialogs in Section~\ref{sec:legal}, before we review the literature on engineering solutions for specifying privacy preferences (with emphasis on the purpose) in Section~\ref{sec:solutions}.

\subsection{Legal Requirements for Consent Dialogs}
\label{sec:legal}

Article~7 of the GDPR describes the requirements of legitimate consent: it needs to be (1)~freely given, (2)~unambiguous, (3)~informed, and (4)~withdrawable at any time~\cite{gdpr}. In the event of a dispute, the data controller must prove that the subject has truly given consent to the processing practices~\cite{eu2018gdpr}.  Specifically, consent must be communicated ``by a statement'' or a ``clear affirmative action''~\cite{gdpr}.
Regarding the clearness of this action, ticking a checkbox on a website is considered an acceptable form, while passiveness or predefined default settings that are not actively declined by the subject do not qualify as consent decisions. The European Court of Justice has just reconfirmed this interpretation~\cite{ecj2019}.

If personal data is collected for more than one purpose, data subjects need to be informed and provided with distinct opt-in choices for every purpose~\cite{gdpr}. Besides stating these principles, the GDPR  intentionally does not specify any design template or rules, and thus leaves the exploration of the design space for consent dialogs to the market participants. 

For the specific case of web cookies, the market has adopted a rough classification of purposes into strictly \emph{necessary} (which presumably do not require consent), \emph{preferences}, \emph{statistics}, and \emph{marketing} (which includes third-party tracking)~\cite{cookiebot2019}\cite[Fig.~4\,(d) of][]{degeling2018we}. This mirrors the approach taken in a user survey by Ackerman et al.\ as early as in 1999~\cite{ackerman1999}. The authors distinguish between cookies for ``customized service'', ``customized advertising'', and ``customized advertising across many websites''. They report a decreasing willingness to agree, from 96\% to 77\% for users classified as ``marginally concerned'' about privacy, and from 43\% to 14\% for so-called ``privacy fundamentalists'' in a sample of 381 US internet users (Fig.~3 of \cite{ackerman1999}).
While the former classification is implemented in popular content management systems, it is by no means the only way of defining purposes. As a result, website operators who can afford specialized lawyers enjoy more freedom in the design of consent dialogs. Others follow common practices in order to minimize legal uncertainty, or to comply with the terms of services of third parties who provide content or code to embed (e.\,g., Google Analytics). The bulk of the burden lands on privacy-aware users, who need to understand and navigate each site's specific model.

\subsection{Technical Solutions for Seeking Consent}
\label{sec:solutions}

Researchers have studied ways to effectively inform users about privacy policies and seek their consent to data processing long before the GDPR.  
For example, a CHI paper from 2001 provides design recommendations for cookie consent dialogs after evaluating design changes of the then popular browsers over time~\cite{millett2001cookies}. The authors criticize browsers in which users had to invest great effort when searching for an alternative to the ``accept all cookies'' default setting.
Consent dialogs are specific forms of privacy notices, a topic so profoundly researched that Schaub et al.~\cite{schaub2015design} saw the need to systematize the literature. According to their proposed taxonomy, the design space can be divided along the dimensions \textit{timing}, \textit{channel}, \textit{modality}, and \textit{control}. In the following, we use this terminology when applicable.

Bergmann~\cite{bergmann2007generic} addresses the problem of complex and incomprehensible privacy choices. The author suggests a design for generic predefined privacy settings (\textit{timing: at setup}) that are summarized in a limited number of categories. He defines four privacy profiles that differ in the acceptance level of transmission and processing of personal data. 
The suggested solution aims at decreasing the user's cognitive effort when selecting suitable privacy settings, but we are not aware of any empirical study to evaluate this approach. 

Pettersson et al.~\cite{pettersson2006ordinary} discuss a similar design with predefined settings. They suggest the adoption of a privacy management system that asks users for consent before transmitting their personal data (\textit{timing: at setup}). Moreover, users' acceptance of data processing practices can be configured in advance and apply to future website visits. However, the authors point out that designing consent forms that are applicable to a large number of different websites is a complex task. It might require compromises on usability as many different settings need to be offered by the system. 
More specifically, Pettersson et al.~\cite{pettersson2005making} propose design paradigms that include suggestions for consent dialogs. Incorporating recommendations by data protection commissioners and legal experts as well as standards established in the PISA project~\cite{borking2001privacy}, the authors present a dialog window with several mandatory and optional fields, an expandable privacy notice, information about the data recipients, and an ``I agree'' button. They also propose methods to overcome habituation by, for instance, using drag-and-drop actions for consent. The authors qualitatively evaluate  their usability tests and find that some users did not fully trust the privacy management system. 

In a follow-up study, Bergman~\cite{bergmann2008testing} empirically explores how to successfully communicate websites' privacy policies to users. Specifically, he compared a conventional interface for online forms to an extended version with additional explanations of privacy information that pops up in tooltips (so-called ``privacy bars'') while filling the form (\textit{timing: just-in-time}). He finds that participants who saw the extended version were significantly more likely to be aware of the policy than the control group. But he did not measure the cost of this sophistication in terms of response time or frictions to usability. Moreover, the screenshots of the extended dialog (Fig.~2 of~\cite{bergmann2008testing}) bears a risk of information overload.  Finally, as the dialog was only tested on desktop computers, it remains unclear how this information can be perceived on small mobile displays.

Tiny displays raise the need for non-interactive forms of privacy preference negotiations.
An established (but meanwhile discontinued) standard for expressing privacy preferences on the web is P3P. The standard lets websites communicate their privacy policies in machine-readable XML format (\textit{modality: machine-readable}). Each XML element represents a component, such as the type of data, the purpose for data collection, and third party recipients~\cite{cranor2003p3p}. A language called Appel has been developed for enabling users to express their privacy preference through predefined rules (\textit{timing: at setup}), so that automated privacy decisions can be based on the user's specific settings~\cite{langheinrich2002appel}. 

A recent approach towards facilitating informed and GDPR-compliant user consent is proposed by Ulbricht and Pallas~\cite{ulbricht2018yappl}. The authors present a privacy preference language, called YaPPL, that is targeted on consent for data practices on the Internet of Things (IoT). For the development, they analyze legal requirements for consent and transform them to technical standards that suit IoT devices (\textit{modality: machine-readable, channel: primary or secondary}). The language is prototypically tested in real-world IoT applications. The authors hope that the underlying approach of YaPPL will also be implemented in IoT applications that do not have to meet the standards of GDPR, but require a technical representation of users' privacy preferences.

Dissatisfied by the observation that many users tend to ignore notices with privacy impact~\cite{vila2004we, grossklags2007empirical}, perceive them as a threat to their privacy~\cite{kulyk2018website}, and have been habituated to ``click away'' consent dialogs~\cite{bohme2010trained}, several researchers investigated how to design more effective privacy notices. For instance, Felt et al.~\cite{felt2012ask} propose design guidelines that aid mobile application developers in appropriately asking for permissions. They find that more than half of all permission requests can be automated while 16\% require consent dialogs. By minimizing the number of runtime consent dialogs, the authors intend to decrease the required user attention. While technical permissions differ from legal purposes in several respects, it is conceivable that similar effects also apply to purposes. 
To our knowledge, this link is still unexplored.

Most closely related to the present work is the concurrent effort by Utz et al.~\cite{utz2019ccs}, which draws on data from a field experiment exploring the design space for cookie banners. Both works share the experimental method, inquiry period (Q1/2019), and language (German). Some of their treatments and findings relate to our research questions. We shall comment on specific similarities and differences where it applies. The most salient differences between our colleagues' and this work are the mode of data collection (field vs lab), the type of cookie notice studied (non-blocking banner vs blocking dialog), the emphasis of the analysis (behavioral traces vs stated attitudes and beliefs), and the context of scientific discovery (inductive vs deductive). Both works leave many questions open, indicating that we are at the beginning of a relevant and potentially fruitful strand of research.
%

The works discussed in this section are selected pieces of the literature. They are representative in that the field focuses on technical and human aspects in many facets, but (with a few exceptions) it largely ignores economic interests~\cite{SABH2015}. In practice, we must expect that businesses use the flexibility in the design of consent dialogs for their own interest by maximizing data disclosure instead of helping users to make privacy-conscious decisions.

\section{Theory}
\label{sec:theory}

User studies integrate better into the body of knowledge (and, arguably, generalize better), if the hypothesized causal links are derived from established theory. Therefore, we revisit relevant theories for explaining the effect of the two characteristic components in the consent dialogs inspiring this work (Fig.~\ref{fig:banner}). Specifically, we review choice proliferation in Section~\ref{sec:choice} to reason about the number of purposes, and social norms in combination with deception in Section~\ref{sec:deception} to predict the effect of the default button. Then, we formulate our hypotheses in Section~\ref{sec:hypotheses}.

\subsection{Choice Proliferation}
\label{sec:choice}

Choice proliferation is a line of research in psychology that analyzes the influence of an increasing number of alternative choices on the human decision-making process. The phenomenon that more options result in negative effects, such as dissatisfaction, has mainly been studied in a marketing context~\cite{scheibehenne2010can,  johnson2012beyond} and is sometimes referred to as ``too much choice'', ``tyranny of choice'', or ``choice overload''. 

As pointed out by Johnson et al.~\cite{johnson2012beyond}, two main aspects have to be considered when evaluating the number of choices offered. On the one hand, a high number of alternatives increases the cognitive load while causing individuals to feel stressed, overwhelmed, and more likely to regret one's decision~\cite{kling2008misperception,cronqvist2004design}. On the other hand, the likelihood that the choice suits the individual's preferences increases when more options are given. Thus, the practical challenge is to find the right balance.

A few works investigate the effect of increasing privacy choices on users' decision making. Korff and B\"ohme~\cite{korff2014too} experimentally study the influence of choice amount and choice structure in the context of privacy preferences on a fictitious business networking website. They find that participants who were confronted with a larger number of privacy settings to chose from were less satisfied with their choice and experienced more regret. The works by Knijnenburg et al.~\cite{knijnenburg2013preference} and Tang et al.~\cite{tang2012implications} investigate the number of privacy choices in the context of mobile location sharing. Both studies find that the structure of presented choices significantly impacts users' tendency to disclose personal data. 
Utz et al.\ vary the number of choices of a cookie banner in their field study, but they neither relate this treatment to choice proliferation nor collect the relevant dependent variables. Since the instrument confounds the number of options with their type (5 categories, with one pre-selected, and 6 vendors; see Fig.~1 (d) and (e) in \cite{utz2019ccs}), it is not easy to interpret the results.
Krasnova et al.\ discuss the effect of an increasing amount of information items in mobile applications' permission requests~\cite{krasnova2013does}. 
The results of their experiment show that users tend to be more concerned if the permission request asks for more information items. This aspect of choice proliferation seems to be specific to privacy, because options in privacy dialogs often remind users of threats. This is rarely the case in the marketing literature on choice proliferation, where the typical study varies the number of forms of a retail product (e.\,g., flavors of jam).

Like almost any social science theory, choice proliferation is not undisputed.
Critics argue that more options can lead to higher satisfaction since one's individual needs can be matched more precisely~\cite{anderson2006long}. Moreover, more choice enables easier comparison of differences, which leads to more confident decision making~\cite{hutchinson2005more}.

Broadly related to the number of options is the number of occasions for privacy decisions. B\"ohme and Grossklags~\cite{bohme2011security} discuss the averse effects of escalating too many decisions to users. They postulate that only the most important decisions should be made by users, so that they do not get habituated to ignore notices as a consequence of too high complexity. Several empirical studies support this interpretation. For example, 
 the null result in an experiment on more or less verbose variants of the well-known consent dialog of Facebook Connect is attributed to habituated ignorance~\cite{egelman2013}. 

Our study connects to the literature on choice proliferation by experimentally varying the number of purposes. We adapt established constructs to measure perceived task difficulty and regret.

\subsection{Deception and Social Norms}
\label{sec:deception}

The concept of deception is often described as being misled due to unfair practices and can occur in many contexts when interests of different parties collide~\cite{johnson2001detecting}. Deception has been studied in several areas such as marketing~\cite{roman2010relational, xiao2011product, nochenson2014online, boush2015deception} and organizational research~\cite{fleming2008escalation, jehn2008perceptions, yoon2018deceiving}. A deceptive practice is being conducted if the targeted individual receives false information that lead to false impressions of a situation. Such false impressions may trigger decisions or opinions that would have been formed in a different way without the deceiving act. 

However, deception is not always based on lying, as it may also comprise purposeful evocation of specific actions by the targeted party; for instance, by increasing the complexity of information, or by making use of behavioral clues or clue patterns. A study by Nochenson and Grossklags~\cite{nochenson2014online} investigates how users of web shops are tricked into falling for post-transaction marketing tactics due to specific design elements in notices. In an experiment, they test the purchasing behavior of more than 500 users and find that above 40\% signed up unintentionally for an extra service with costs. The authors find that opt-in and opt-out default buttons significantly impact the users' tendency to fall for the trick. 

Citing usability guidelines~\cite{shneiderman2006research}, B\"ohme and K\"opsell~\cite{bohme2010trained} underline that the default option should include the most frequently selected settings so that inexperienced users can be assisted by the decision of the majority. In this sense, default buttons can be interpreted as a descriptive social norm. However, as highlighted in a study on default privacy settings on social media websites, the preset or default options are often very disclosing and might not reflect the majority of users' privacy preferences~\cite{watson2015mapping}. It seems that the default button has mutated from a usability tool that improves efficiency when selecting the typical choice to a strategic tool that supports the interests of the system designer. 

For several decades, scholars in the behavioral sciences have identified and quantified cognitive and social effects, some of which cause successful persuasion or deception~\cite{hovland1953yale}. A shared objective in these disciplines was to isolate effects, which required substantial effort given that stimuli to human subjects often confound many factors. By contrast, the recent literature that criticizes the deliberate exploitation of these biases in favor of the designer typically looks at bundles of features as they appear in practice~\cite{gray2018darkpattern}. The term ``dark pattern''~\cite{brignull2018}, coined in 2010, classifies designs 
that trick users into making decisions they do not mean to make. 
B\"osch et al.~\cite{bosch2016tales} were among the first to systematize dark patterns commonly adopted for privacy invasions. 
For instance, users typically do not read privacy notices completely~\cite{mcdonald2008cost} and often intuitively accept the presented conditions. This behavior can be exploited by hiding undesirable terms in privacy notices. 
Mathur et al.~\cite{mathur2019dark} structure common characteristics of dark patterns along five dimensions: (1) \textit{asymmmetric} (unequal emphasis or obstacles for specific choices), (2) \textit{covert} (hidden interface design choices), \textit{deceptive} (induce false beliefs), (3) \textit{hides information} (obscure or delay the communication of relevant information), and (5) \textit{restrictive} (limitation of choices). The authors specifically name cookie consent dialogs which make use of a highlighted ``accept'' button as an example for the \textit{asymmmetric} dimension.

In the context of the GDPR, one could argue that tactics involving increased complexity, hidden information, or unwanted default settings---if effective---violate the requirements for clear and informed consent. Our study adds empirical evidence on the effectiveness of these tactics in the specific context. We vary the presence of a potentially misleading default button and measure perceived deception, unlike the wealth of studies that quantify this bias by merely observing the behavioral reaction to default buttons.\footnote{The default effect is in the order of 5 \%-pts.\ for a consent dialog where about one of two participants agrees~\cite{bohme2010trained}.} Since decisions in the privacy context often involve high cognitive load, 
we devise a combined (but not confounded) experiment with choice proliferation. This allows us to interpret perceived difficulty and response time---both proxies for cognitive load---in relation to perceived deception.

\subsection{Hypotheses}
\label{sec:hypotheses}

Against the backdrop of the features in consent dialogs used by popular websites and the underlying theoretical considerations, we postulate four hypotheses:

\begin{description}
	\item[H1] If consent dialogs include a highlighted default button that selects all purposes, users effectively consent to more purposes than without this button.
	\item[H2] If consent dialogs include a highlighted default button that selects all purposes, users \\
	\textbf{(a)} regret their decision more and\\
	\textbf{(b)}\,perceive the website as more deceptive\\than without this button,
	after being informed about the purposes they effectively consented to.
	\item[H3] If consent dialogs present multiple purposes, users require more effort than for dialogs with a single purpose, as indicated by longer response times.
	\item[H4] If consent dialogs include multiple purposes, users perceive the task as more difficult than reacting to dialogs with a single purpose.
\end{description}

In the hypotheses and the following, we shall use the term ``effective consent'' to refer to the consent statement recorded by the website, independent of whether this corresponds the user's true intention.

\section{Method}
\label{sec:method}


To test the proposed hypotheses, we conducted a controlled experiment. We describe and justify the instrument in Section~\ref{sec:instrument}, then report from our pretests (Section~\ref{sec:pretest}) and the survey administration (Section~\ref{sec:adiminstration}). Ethical considerations are discussed in Section~\ref{sec:ethics}. Descriptive statistics are presented in Section~\ref{sec:descriptives}.

\subsection{Instrument}
\label{sec:instrument}
\begin{figure*}[tbp]
	\begin{center}
		\begin{tabular}{ccc}
			\includegraphics[width=5.5cm]{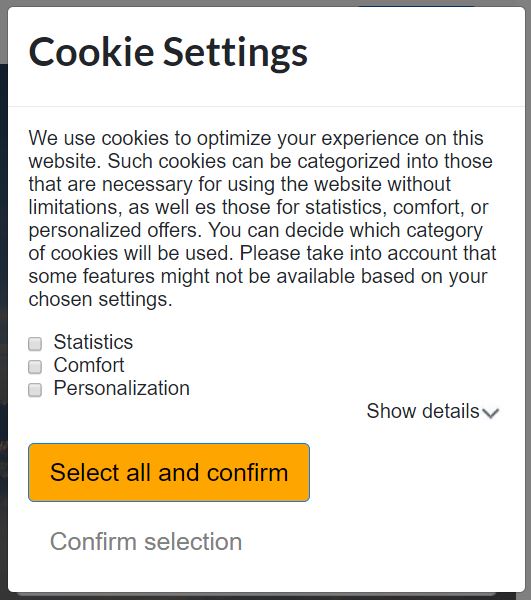} & 
			\includegraphics[width=5.5cm]{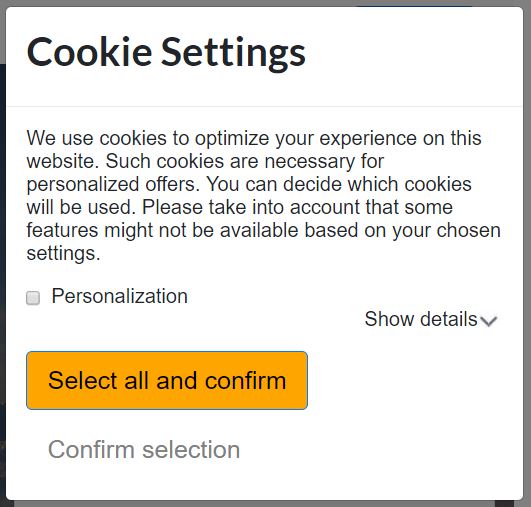} & 
			\includegraphics[width=5.5cm]{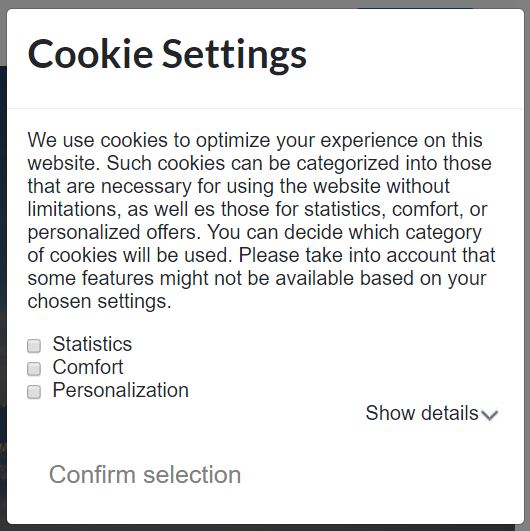} \\
			Deception (T1)  & Reduced choice (T2)   & Control \\                             
		\end{tabular}
	\end{center}
	\caption{\label{fig:banner_versions} Variations of consent dialogs shown in the study. All dialogs are blocking. The  participants saw German versions (see Fig.~\ref{fig:banner_versions_german}).}
\end{figure*}

The survey instrument has two main components: a functional mock-up website offering flight search, and an exit questionnaire. As experimental factor, the mock-up randomly presents the user one of the three consent dialogs depicted in Figure~\ref{fig:banner_versions}. When categorizing these dialogs along the dimensions proposed by Schaub et al.~\cite{schaub2015design}, they constitute privacy notices which appear at setup (\textit{timing}), in the primary \textit{channel}, as visual pop-ups (\textit{modality}) that include a blocking \textit{control}.
%
We copied the three purposes (statistics, comfort, personalization) from the airline website in verbatim in order to maximize external validity, noting that they differ from the convention discussed in Section~\ref{sec:solutions}. Users could learn more about the purposes by clicking on a small roll-down button labelled ``show details'' (see screenshot in Fig.~\ref{fig:details_german} in the Appendix). Accordingly, \emph{comfort} corresponds to \emph{prefereces}, and \emph{personalization} to \emph{marketing}, however without an indication whether this includes third-party tracking.

The treatment of the first group (T1) is a deceptive dialog, which closely resembled the one we saw on the German airline website (cf. Figure~\ref{fig:website_german}). It contains an explanation text about different cookie settings, three selectable purposes with initially unchecked checkboxes, an expandable part providing more details about the categories, and two buttons. 
The first button with the text ``Select all and confirm'' stands out due to its yellow color. The second button is colorless and says ``Confirm selection'' in gray font. If the yellow button is clicked, the user (effectively) consents to all three purposes, regardless of which boxes are checked. In contrast, a click on the second button only confirms the settings that have actively been selected by the user. 

The second treatment (T2) differs in the reduction of selectable categories. Specifically, it only includes the \emph{personalization} purpose. Pretests have shown that personalization is perceived as the most sensitive purpose, thus we deemed it plausible to make this purpose optional. 
We could confirm this post-hoc: only 23\% of the users in the control group consent to \emph{personalization}, versus 35\% for \emph{comfort} and 46\% for \emph{statistics}. The results of Utz et al.~\cite{utz2019ccs} corroborate this further.\footnote{See Fig.~5\,(1a) of~\cite{utz2019ccs}, although the precision is low and the baseline not comparable.}
Arguable, the T2 dialog appears somewhat artificial, but it was the best way we could think of reducing the number of choices without changing the dialog to a yes/no question. 
We could not spot any indication that users perceived this dialog as odd in the responses to an open-ended question in the exit survey.

In contrast to the two treatments, the control group did not see a highlighted default button. The control dialog offers the same three purposes as observed in reality. We refrained from presenting a version with one purpose and no default button for the lack of hypotheses on potential interaction effects, and to increase the number of subjects in the interesting three groups. Therefore, our study technically combines two $1 \times 2$ experiments with one overlapping group rather than realizing a complete $2 \times 2$ design. 

We decided against additional treatments with opt-out (i.\,e., where purposes are pre-selected) because they are almost certainly not compliant with the GDPR~\cite{ecj2019}. For the same reasons, we see little prospect for non-blocking cookie banners if the website to some extent depends on consent as the legal basis to process personal data. For comparison, Utz et al.~\cite{utz2019ccs} test two opt-out conditions in their field study of non-blocking banners.

The actual flight search website has a simplistic design and only contains text fields and date selectors for the search input. To increase realism, some ``special offers'' for specific destinations are depicted next to a photo of the respective city. These measures were intended to draw the focus away from the cookie dialog. The participants' interaction on the website is captured and continuously transmitted to our server. This allows us to analyze response time, click trajectories, and possible dropouts post-hoc.

We measure the participants' perceptions of the website in an exit questionnaire. At first, participants are asked to freely list positive and negative aspects of the website. Thereafter, they should recall their chosen cookie settings in the dialog; first in free-text form and followed by closed questions. Besides general questions on the cookie dialog, four established constructs are measured through multi-item scales. Such scales are common in psychometrics to attenuate the measurement error of individual items. All construct items are reported in Table~\ref{tab:construct_items}. Answers were collected on 5-point rating scales with semantic anchors ``strongly disagree'' (1) and ``strongly agree'' (5). Perceived deception (PDE) is assessed using three (of originally four) items by Román~\cite{roman2010relational}, adapted to the context of our study.\footnote{The fourth item was dropped because it was too specific to the domain of online shopping.} Additionally, we measure perceived difficulty (PDI), privacy attitudes (PA), and regret (RE). RE is measured twice in the questionnaire: before and after reminding the participants of their effective cookies settings.

\begin{table}[t!]
	\centering
	\caption{Constructs and corresponding items.}
	\label{tab:construct_items}
	\begin{tabular}{p{1.4cm}p{6.3cm}}
		{\textbf{Item}} & {\textbf{Item text (translated from German)}} \\ \midrule
		\multicolumn{2}{@{}p{7.5cm}}{\textit{Perceived Deception (PDE)}}  \\
		\hspace{0.18cm}PDE1 & \PDEa   \\ 
		\hspace{0.18cm}PDE2 & \PDEb   \\
		\hspace{0.18cm}PDE3 & \PDEc   \\[1ex]
		\multicolumn{2}{@{}p{7.5cm}}{\textit{Perceived Difficulty (PDI)}}\\
		\hspace{0.18cm}PDI1 & \PDIa \\ 
		\hspace{0.18cm}PDI2 & \PDIb \\ 
		\hspace{0.18cm}PDI3$\leftrightarrow$ & \PDIc \\[1ex]
		\multicolumn{2}{@{}p{7.5cm}}{\textit{Regret (RE)}}  \\ 
		\hspace{0.18cm}RE1 & \RGa \\ 
		\hspace{0.18cm}RE2 & \RGb \\ 
		\hspace{0.18cm}RE3$\leftrightarrow$ & \RGc \\[1ex] 	
		\multicolumn{2}{@{}p{7.5cm}}{\textit{Privacy Attitudes (PA)}}  \\ 
		\hspace{0.18cm}PA1 & \PAa \\ 
		\hspace{0.18cm}PA2 & \PAb \\ 
		\hspace{0.18cm}PA3 & \PAc \\
	\end{tabular}\\[1ex]
	\small
	Items marked with `$\leftrightarrow$' use inverted scales.
\end{table}

\subsection{Pretests}
\label{sec:pretest}
Two pretests were conducted in order to assess the clarity of the instructions and survey questions. First, we carried out two one-on-one tests using \textit{verbal probing} and \textit{think aloud} techniques. Specifically, test subjects were asked to express their thought process and potential obstacles while going through the survey. Since we found that it is confusing to first open a link with a cookie dialog, and then receive the flight search task, we decided to rearrange the instructions. This way, the participants are even more focused on flights than on cookies before visiting the website. 

In order to simulate the actual survey environment in a lecture hall, the second pretest was conducted with 20 Austrian undergraduate students in a computer lab. This way, we were able to estimate the required time for each survey step. During the test, we observed that several test subjects glanced at their neighbors' screens and talked to one another while completing the survey. As this behavior might reduce the data quality, we added the appeal to work quietly and by oneself to the instructions. Additionally, we found and fixed a bug concerning the collection of timestamps.

\subsection{Survey Administration}
\label{sec:adiminstration}

\begin{figure*}[t]
	\begin{center}
		\scalebox{0.9}{
		\input{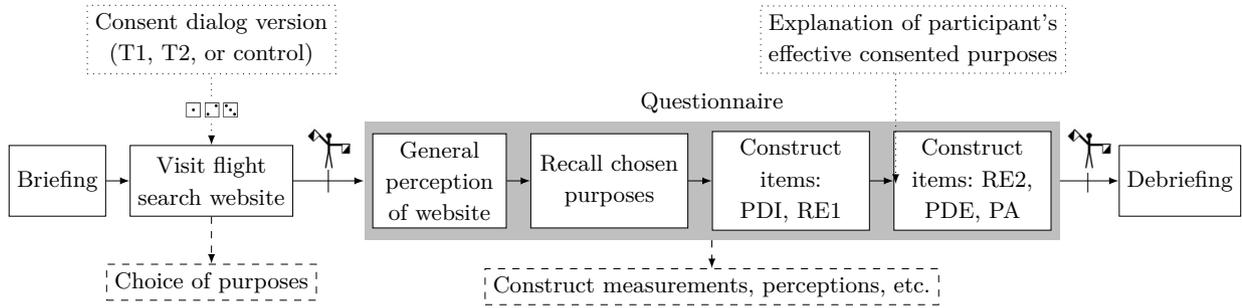}
	}
	\end{center}
	\caption{\label{fig:study_steps} Visualization of the study process with treatments (dotted boxes) and measurements (dashed boxes). 
	Semaphores denote synchronization points: all participants in the classroom proceed to the next step simultaneously.}
\end{figure*}

The data collection took place on two days in January 2019 at the University of Innsbruck in Austria and the University of Münster in Germany. All parts of the instrument and the written and spoken instructions were provided in the local language (German). We report the original wording of scale items and selected screenshots in the appendix (Section~\ref{sec:appendix}) to facilitate error analysis and possible replication studies. The survey was administered at the beginning of lectures attended by undergraduate computer science students, mainly in the first year. Figure~\ref{fig:study_steps} summarizes the steps of the data collection, along with the order of the measured constructs.

During the briefing, we informed the participants that their data will be held confidential and cannot be linked to their identity. We also pointed out the voluntary nature of participating in the study and asked them to conscientiously follow the instructions without interacting with one another. We communicated that the scope of the study is about the user experience of flight search websites, without mentioning the focus on cookie notices or privacy. 

In the second step, participants were given the task to search for a flight with a specific departure, destination, and time. Then, we provided the link to the flight search website that is described above. When visiting the link, one of the three cookie dialogs depicted in Figure~\ref{fig:banner_versions} was randomly assigned to each participant. After reacting to the dialog and entering the flight search, a modal window appeared, which asked to wait for further instructions. 

When the vast majority had reached this step, a key combination for opening the questionnaire was displayed on the lecture hall's main projector. This way, participants started answering the questions almost simultaneously. It took the average participant 6' 34'' to complete the questionnaire while $90\%$ were done after 9' 06''.  

In the debriefing, we informed the participants about the topic of the study and showed them screenshots of cookie notices used by real websites. 
We ran the classroom experiment exactly once at each university, one in Austria and one in Germany, thereby minimizing the likelihood that earlier participants could tell later participants the true purpose of the study.

\subsection{Research Ethics}
\label{sec:ethics}
In fulfillment of approved ethical standards, we clearly communicated that participation is voluntary and anonymous. Respondents could skip questions they did not want to answer. The search task itself and the surrounding stimulus material was chosen to not raise emotions or strong feelings.  
Independent of the participants' selected cookie settings, we did not store cookies and only transmitted data to our servers that are relevant for the research purpose. 

In not revealing the true purpose of our study right away, we applied deception ourselves as part of the research method. This is common practice and was in accordance with the ethical oversight bodies at all universities involved. The practice is deemed acceptable in particular because of the low probability of causing harm and the fact that we revealed the purpose of our study in the debriefing, where we also provided contact information and offered the communication of results.

The experiment caused an opportunity cost of 10 minutes lost lecture time for everyone in the room, including about 8 students per session (15 altogether) who did not participate in the experiment. To minimize the harm, we chose the beginning of a Q\&A session that had not used the entire allocated time in the previous years. Moreover, the main reason why students did not participate was that they arrived late in class.

\subsection{Descriptive Statistics}
\label{sec:descriptives}

\begin{table}[b!]
	\centering
	\caption{Descriptive statistics.}
	\label{tab:demo}
	\begin{tabular}{p{4.4cm} r r }
		{ \textbf{Item}} & { \textbf{Number}} & \textbf{Fraction} 
		\\ 		\midrule 
    \hspace{0.4cm}All&150& 	$100.0\%$ 
    \\[1ex] 
    
    \textit{{Group}} & &\\
    \hspace{0.4cm} T1 & $50$ & $33.3\%$\\
    \hspace{0.4cm} T2 & $48$ &  $32.0\%$ \\
    \hspace{0.4cm} Control & $52$ &  $34.7\%$ \\[1ex]
    
    \textit{{Location of the university}} & &\\
    \hspace{0.4cm} Austria & $90$ & $60.0\%$ \\
    \hspace{0.4cm} Germany & $60$ &  $40.0\%$ \\[1ex]
    
    \textit{{Screen width}} & &\\
    \hspace{0.4cm}$<$500px & $64$ & $42.7\%$  \\
    \hspace{0.4cm}500px--1000px & $7$ &  $4.7\%$ \\
    \hspace{0.4cm}$>$1000px & $79$ &  $52.7\%$ \\[1ex]
    
    \textit{{Browser}} & &\\
    \hspace{0.4cm} Chrome & $81$ & $54.0\%$ \\
    \hspace{0.4cm} Firefox & $30$ & $20.0\%$ \\
    \hspace{0.4cm} Safari & $29$ & $19.3\%$ \\
    \hspace{0.4cm} Other & $10$ & $6.7\%$ \\[1ex]
    
    \textit{{Knowledge about cookies}} & &\\
    \hspace{0.4cm} Self-reported knowledge & $121$ & $80.7\%$ \\
    \hspace{0.4cm} Correctly described cookies & $102$ & $68.0\%$  \\[1ex]
    
    \textit{{Privacy measures (self-report)}} & &\\
    \hspace{0.4cm} Regularly deletes cookies & $64$ & $42.7\%$  \\
    \hspace{0.4cm} Has cookies disabled & $36$ & $24.0\%$\\
    \hspace{0.4cm} Uses ad-blocker & $113$ & $75.3\%$\\
    \hspace{0.4cm} Uses anti-virus software & $80$ & $53.3\%$\\
		
	\end{tabular}
\end{table}

Table~\ref{tab:demo} reports descriptive statistics of our sample. In total, 164 students took part in the study whereof 158 completed the survey. We deleted 8 records due to more than four missing responses on critical construct items. The remaining 13 records with missing answers contained a total of 20 missing values, which were replaced by the mean of the observed item score. 
Consequently, the analysis uses 150 valid cases. As a consequence of the convenience sample, the ratio of female participants was below 20\%, which is typical for German-speaking computer science undergraduates.

Even though we asked participants to use their laptops for completing the survey, we allowed those without one to chose another device they had at hand. While 42.7\% followed the survey instructions on a screen width below 500 pixels (i.\,e., likely smartphones), 
52.7\% had a screen width above 1000 pixels (i.\,e., likely laptops). Most participants opened the website on Chrome (54.0\%); others used Safari (19.3\%) or Firefox (20.0\%).  We did not observe noteworthy differences in results between device types, browsers, or locations and thus refrain from reporting breakdowns in the following.

By asking whether participants know what browser cookies are, we find that 68.0\% are able to provide a correct explanation. Only 12.7\% claim to know what cookies are, but provided either no explanation or an incorrect one. The remaining 19.3\% stated that they have no knowledge about cookies.

On average, it took participants 11.8 seconds to respond to the cookie dialog (median: 7.3''). Only 8.7\% expanded the dialog by clicking on ``Show details'', and 3.3\% revised their initial choice by unselecting at least one purpose. In total, 41.3\% did not consent to any cookie purpose, while 30.7\% accepted all purposes offered. Of all participants in the two treatment groups ($n=98$), 56.1\% clicked on the default button, which results in accepting all purposes regardless of which purposes were actively selected. Of these participants, 34.5\% ($n=19$) still selected at least one purpose. 

To evaluate the construct reliability of PDE, PDI, RE und PA, we examine internal consistency by calculating Cronbach's $\alpha$. As shown in Table~\ref{tab:construct}, each constructs'  Cronbach's $\alpha$ value lies above 0.7, indicating that they are sufficiently consistent~\cite{bland1997cronbach} and thus suitable for further analysis. We also check if the construct scores are sufficiently close to a normal distribution to justify the use of parametric inference statistics. Table~\ref{tab:normality} shows Q-Q plots for all constructs and reports the results of Kolmogorov--Smirnov (KS) tests for normality. Given that the deviations from normality are visibly caused by the range limits only, no KS-test rejects the null hypothesis at the 1\% level, and the way we compute the scores cannot produce any outliers, we deem it safe to report hypothesis tests with parametric $t$-tests. To err on the side of caution, we report $p$-values for the two-sided test although all our hypotheses are directed.

\begin{table}[]
	
	\caption{Construct reliability.}
	\label{tab:construct}
	\begin{center}
	\begin{tabular}{l r r r r r}
		{\textbf{Construct}} & { \textbf{Cronbach's $\alpha$}} & \textbf{Mean} & \textbf{Median}   & \textbf{SD}  
		\\ 		\midrule 
		
    PDE        & $0.79$ & $3.52$ & $3.67$ & $1.13$ \\
    PDI        & $0.73$ & $2.82$ & $2.67$ & $1.17$ \\
    RE-before  & $0.83$ & $2.39$ & $2.33$ & $1.14$ \\
    RE-after   & $0.74$ & $2.62$ & $2.33$ & $1.20$ \\
    PA         & $0.80$ & $3.61$ & $3.67$ & $0.91$ \\
    
	\end{tabular}\\[1ex]
	\small Each construct has 3 items.\\
\end{center}
\end{table}

\begin{table}[]
	
	\caption{Normality of construct distributions.}
	\label{tab:normality}
	
	\begin{center}
	\begin{tabular}{c@{~~}c@{~~}c@{~~}c@{~~}c}
	\textbf{PDE} & \textbf{PDI} & \textbf{RE-before} & \textbf{RE-after} & \textbf{PA} \\
	\midrule
	\multicolumn{5}{@{}l}{Normal Q-Q plots for the range $\pm3$ SD} \\
	\tikz{\node [inner sep=0pt] (A) {\includegraphics[width=14mm]{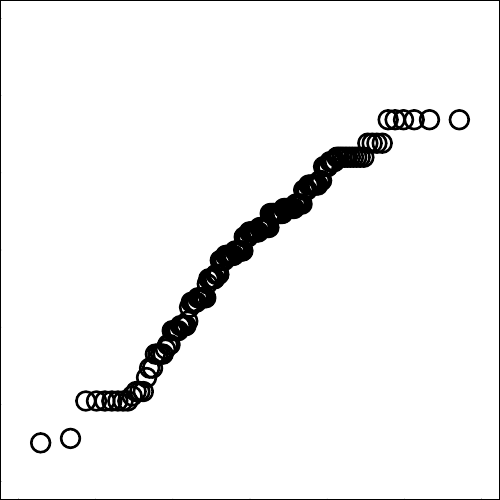}}; 
		\draw (A.south west)--(A.north east);
		}
		&
	\tikz{\node [inner sep=0pt] (A) {\includegraphics[width=14mm]{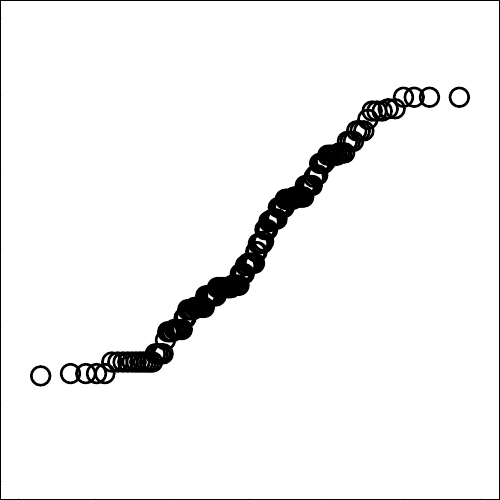}}; 
		\draw (A.south west)--(A.north east);
		}
		&
	\tikz{\node [inner sep=0pt] (A) {\includegraphics[width=14mm]{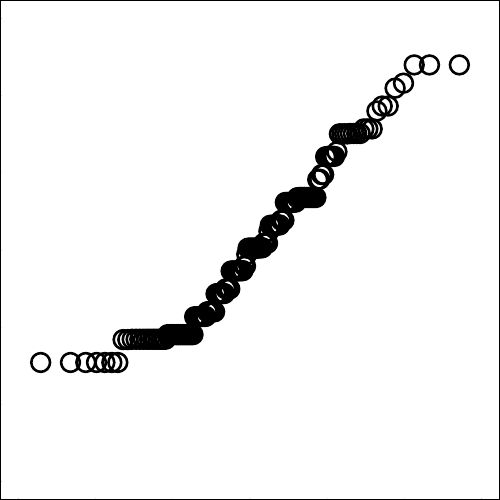}}; 
		\draw (A.south west)--(A.north east);
		}
		&
	\tikz{\node [inner sep=0pt] (A) {\includegraphics[width=14mm]{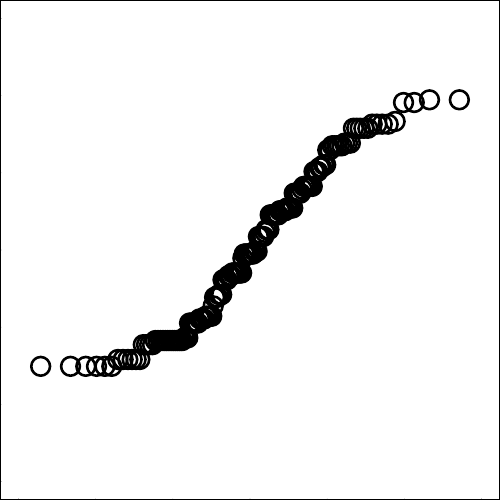}}; 
		\draw (A.south west)--(A.north east);
		}
		&
	\tikz{\node [inner sep=0pt] (A) {\includegraphics[width=14mm]{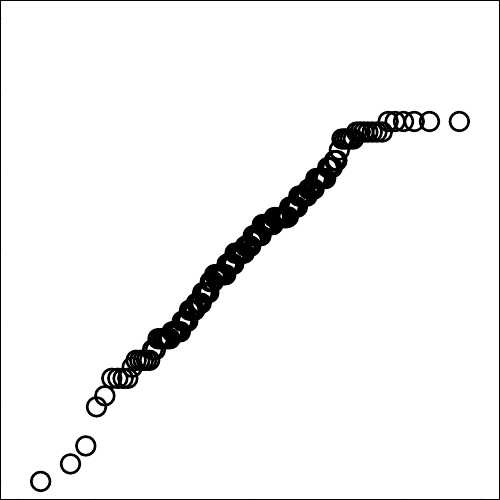}}; 
		\draw (A.south west)--(A.north east);
		}
	\\
	\multicolumn{5}{@{}l}{One-sample Kolmogorov--Smirnov tests for normality} \\
	$ p = 0.10$ 
	&
	$ p = 0.02$ 
	&
	$ p = 0.04$ 
	&
	$ p = 0.03$ 
	&
	$ p = 0.47$ 
	\\
	\end{tabular}
	\end{center}

\end{table}

\section{Results}
\label{sec:results}

We begin with the deductive hypothesis tests, before we investigate additional aspects in a quantitative explorative way (Section~\ref{ssec:posthoc}).

\subsection{Hypothesis Tests}

We test \textbf{H1} by analyzing whether the deception group and the control group differ in the number of purposes they effectively agreed to. To do so, we assign a score from 0 (no purposes, by clicking on ``Confirm selection'' without checking any box) to 3 (all purposes by either checking all boxes and then clicking any button, or by clicking the highlighted default button). 
Participants who saw the deceptive dialog effectively consented to more purposes. Table~\ref{tab:hypotheses_tests} presents the score values by treatment and control group. Since the score is a count variable, we use the non-parametric Kruskal--Wallis (KW) test, which indicates as strongly significant effect  ($\chi^2(1)=7.2$, $p<0.01$)\unskip in the hypothesized direction. \textbf{This supports H1.}
The effect of the deceptive default button on agreeing to no or all purposes is in the order of 20 \%-pts., about four times larger than the plain default effect reported for an application consent dialog in~\cite{bohme2010trained}.
We additionally check if participants in the deception group are more likely to consent to all three, instead of two or less purposes, than the control group. A Chi-squared test ($\chi^2(1)=9.05$, $p<0.005$)\unskip reveals a highly significant difference.

To test \textbf{H2a}, we compare the measurements of regret before (RE-before) and after (RE-after) the participants got informed about the purposes they effectively consented to. Results of the paired $t$-test reveal a significant difference between the before/after states for the deception group ($t(49)=2.81$, $p<0.01$, $d=0.40$)\unskip. \textbf{This supports H2a.} The  difference in the control group is not significant ($t(51)=1.63$, $p=0.11$, $d=0.23$). Thus, we can attribute the regret to the misinformation caused by the deceptive design.

When analyzing perceived deception (PDE) for testing \textbf{H2b}, a notable difference between groups can be found (Figure~\ref{fig:boxplot_deception}). The $t$-test shows that PDE of the deception group is significantly higher than of the control group ($t(96.8279)=2.24$, $p<0.05$, $d =0.44$)\unskip. Therefore, \textbf{H2b is also supported.} 

Drilling down into the findings on H2a and H2b, we analyze if participants within the deception group who clicked on the deceptive default button perceive even more regret and deception after being informed about the consequence of their response. Indeed, our measurements of RE-after are significantly higher for those who clicked the default button compared to all other participants in T1 ($t(43.962)=-3.82$, $p<0.0005$)\unskip. 
However, no significant differences for perceived deception can be found ($t(47.73422)=0.64$, $p=0.64$)\unskip. These two results can be explained by the presence of smart participants who debunk the default button as deceptive and do not fall for it. They have less to regret than those who only understand the button's effect after the fact.

To test \textbf{H3}, we investigate the time needed to complete the consent dialogs. The measurement starts when the cookie dialog appears and ends when the participant clicks a button. This measure reflects the effort required for responding to the dialog.
As shown in Figure~\ref{fig:boxplot_time}, participants in the group with reduced choice spent on average five seconds less on their response than those who were presented with three purposes. 
The difference in medians shows the same trend, albeit less pronounced due to the skewed distribution. We choose non-parametric statistics to account for this fact.
When only comparing the deception and reduced choice group, the KW-test reveals that the difference ($\chi^2(1)=8.89$, $p<0.005$)\unskip is highly significant, which \textbf{supports H3.} We also find a significant difference between the reduced choice group and the control group ($\chi^2(1)=9.73$, $p<0.005$). The difference between the deception and control group, which offer the same number of purposes, is not significant ($\chi^2(1)=0.17$, $p=0.68$). The results indicate that the number of purposes is positively associated with cognitive load, even if the number of options is way below Miller's ``magic seven''~\cite{miller1956}.

Regarding \textbf{H4}, we interpret perceived difficulty (PDI) as a measure of dissatisfaction. Specifically, we test the difference between the two treatment groups in order to show whether the number of purposes in the consent dialog affect the participants' perceptions. Since the $t$-test results in no significant difference ($t( 95.99)= 0.16 $, $p=0.88$, $d=0.07$), \textbf{H4 must be rejected.} Unrelated to our hypotheses, we also tested for differences in PDI between the control group and T1, respectively, T2. No test result was even close to statistical significance.


\begin{table}[t!]
	\caption{\label{tab:hypotheses_tests} Overview of results by treatment group.}
	\begin{center}
		\begin{tabular}{p{1.6cm} r rr  p{0.2cm} p{0.2cm}}
			& \multicolumn{3}{c}{\textbf{Group}} & & \\ 
			\cmidrule{2-4} 
			\textbf{Dependent}& \multicolumn{1}{c}{\textbf{T1}}& \multicolumn{1}{c}{\textbf{T2}}& \multicolumn{1}{c}{\textbf{Control}}\\
			\vspace{-0.3cm}
			\textbf{variable} & {\small($n=50$)} & {\small($n=48$)} &  {\small($n=52$)}& \multicolumn{2}{c}{\textbf{Test}}     \\ 
			\midrule 

			\multicolumn{4}{@{}l}{\textit{Number of effectively consented purposes}} & \multicolumn{2}{c}{KW-test} \\
			\cmidrule{5-6}
			0 & $\textbf{32.0\%}$ & $41.7\%$ & $\textbf{50.0\%}$ & &\\ 
			1 & $\textbf{12.0\%}$ & $58.3\%$   & $\textbf{19.2\%}$ & &\\ 
			2 & $\textbf{2.0\%}$ & --          & $ \textbf{7.7\%}$ &  &\\
			3 & $\textbf{54.0\%}$ & --         & $\textbf{23.1\%}$ &  **& H1\\
			\cmidrule{2-4} 
			& $100.0\%$     & $100.0\%$    & $100.0\%$ &  & \\\\
			
			\multicolumn{4}{@{}l}{\textit{Construct means}} & \multicolumn{2}{c}{$t$-test} \\
			\cmidrule{5-6}
			PDE & \textbf{3.68} & 3.74 & \textbf{3.15} & * & H2b  \\
			PDI  &  \textbf{2.87} & \textbf{2.90} & 2.69  & n.s. & H4 \\
			RE-before & \textbf{2.30} & 2.64 & \textbf{2.23} &  n.s. & \\ 
			RE-after & \textbf{2.69} & 2.79 & \textbf{2.39} &  n.s. & \\
			PA & 3.50 & 3.70 &  3.62 & \\[1ex] 
			\multicolumn{6}{l}{\textit{-- '' -- \quad (only subjects who clicked the default button)}} \\
			& {\small($n=27$)} & {\small($n=28$)} &&&\\
			\quad PDI & 2.78 & 2.72 && &\\
			\quad RE-after & 3.20 & 3.48 & --  &&\\
			\quad PA & 3.32 & 3.56 &  -- &  &\\\\ 
			
			\multicolumn{4}{@{}l}{\textit{RE-after minus RE-before \color{gray}(all subjects)}} & \multicolumn{2}{r@{}}{paired $t$-test} \\
			\cmidrule{5-6}
			& \textbf{0.39} & 0.15 & 0.16 & **  & H2a  \\\\
			
			\multicolumn{4}{@{}l}{\textit{Response time for consent dialog (seconds)}} & \multicolumn{2}{c}{KW-test} \\
			\cmidrule{5-6}
			Median & \textbf{5.36} & \textbf{3.16} & 7.24 & ** & H3  \\
			Mean & 10.95 & 5.41 & 11.62 &  &  \\\\
			
			\multicolumn{4}{@{}l}{\textit{Correct recall of effective purposes}} & \multicolumn{2}{c}{$\chi^2$-test} \\
			\cmidrule{5-6}
			& $\textbf{73.5}\%$ & $\textbf{60.9\%}$ & $\textbf{90.0\%} $   & ** &\\ 
			\multicolumn{6}{l}{\textit{-- '' -- \quad (only subjects who clicked the default button)}} \\
			& {\small($n=27$)} & {\small($n=28$)} &&&\\
			& $55.6\%$ & $ 53.6\%$ & --    & &\\ 
			
			\bottomrule
			\multicolumn{6}{l}{Legend: * $p<0.05$, ** $p<0.01$, n.s. not significant.}  \\
			\multicolumn{6}{l}{The test results refer to the bold values in the same row.}  \\
			
		\end{tabular}
	\end{center}
\end{table}

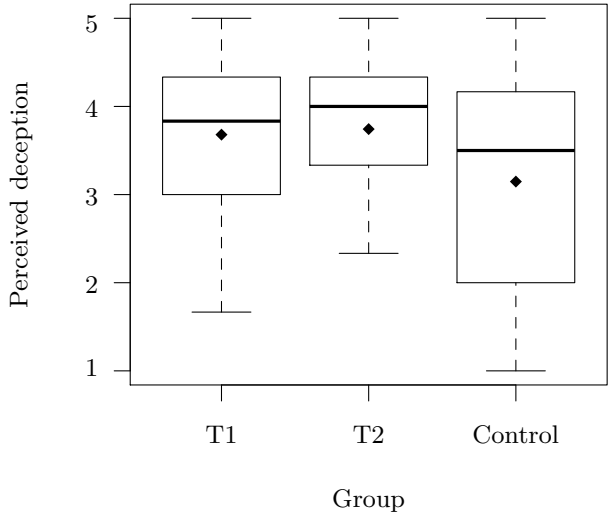
\begin{figure}[t!]
	\begin{center}
		\vspace{-10ex}
\begin{tikzpicture}[x=1pt,y=1pt]
\definecolor{fillColor}{RGB}{1,1,1}
\path[use as bounding box,fill=fillColor,fill opacity=0.00] (0,0) rectangle (252.94,252.94);
\begin{scope}
\path[clip] ( 49.20, 61.20) rectangle (227.75,203.75);
\definecolor{drawColor}{RGB}{0,0,0}

\path[draw=drawColor,line width= 1.2pt,line join=round] ( 61.32,159.97) -- (105.41,159.97);

\path[draw=drawColor,line width= 0.4pt,dash pattern=on 4pt off 4pt ,line join=round,line cap=round] ( 83.37, 88.48) -- ( 83.37,132.47);

\path[draw=drawColor,line width= 0.4pt,dash pattern=on 4pt off 4pt ,line join=round,line cap=round] ( 83.37,198.47) -- ( 83.37,176.47);

\path[draw=drawColor,line width= 0.4pt,line join=round,line cap=round] ( 72.34, 88.48) -- ( 94.39, 88.48);

\path[draw=drawColor,line width= 0.4pt,line join=round,line cap=round] ( 72.34,198.47) -- ( 94.39,198.47);

\path[draw=drawColor,line width= 0.4pt,line join=round,line cap=round] ( 61.32,132.47) --
	(105.41,132.47) --
	(105.41,176.47) --
	( 61.32,176.47) --
	( 61.32,132.47);

\path[draw=drawColor,line width= 1.2pt,line join=round] (116.43,165.47) -- (160.52,165.47);

\path[draw=drawColor,line width= 0.4pt,dash pattern=on 4pt off 4pt ,line join=round,line cap=round] (138.47,110.47) -- (138.47,143.47);

\path[draw=drawColor,line width= 0.4pt,dash pattern=on 4pt off 4pt ,line join=round,line cap=round] (138.47,198.47) -- (138.47,176.47);

\path[draw=drawColor,line width= 0.4pt,line join=round,line cap=round] (127.45,110.47) -- (149.49,110.47);

\path[draw=drawColor,line width= 0.4pt,line join=round,line cap=round] (127.45,198.47) -- (149.49,198.47);

\path[draw=drawColor,line width= 0.4pt,line join=round,line cap=round] (116.43,143.47) --
	(160.52,143.47) --
	(160.52,176.47) --
	(116.43,176.47) --
	(116.43,143.47);

\path[draw=drawColor,line width= 1.2pt,line join=round] (171.54,148.97) -- (215.62,148.97);

\path[draw=drawColor,line width= 0.4pt,dash pattern=on 4pt off 4pt ,line join=round,line cap=round] (193.58, 66.48) -- (193.58, 99.48);

\path[draw=drawColor,line width= 0.4pt,dash pattern=on 4pt off 4pt ,line join=round,line cap=round] (193.58,198.47) -- (193.58,170.97);

\path[draw=drawColor,line width= 0.4pt,line join=round,line cap=round] (182.56, 66.48) -- (204.60, 66.48);

\path[draw=drawColor,line width= 0.4pt,line join=round,line cap=round] (182.56,198.47) -- (204.60,198.47);

\path[draw=drawColor,line width= 0.4pt,line join=round,line cap=round] (171.54, 99.48) --
	(215.62, 99.48) --
	(215.62,170.97) --
	(171.54,170.97) --
	(171.54, 99.48);
\end{scope}
\begin{scope}
\path[clip] (  0.00,  0.00) rectangle (252.94,252.94);
\definecolor{drawColor}{RGB}{0,0,0}

\path[draw=drawColor,line width= 0.4pt,line join=round,line cap=round] ( 83.37, 61.20) -- (193.58, 61.20);

\path[draw=drawColor,line width= 0.4pt,line join=round,line cap=round] ( 83.37, 61.20) -- ( 83.37, 55.20);

\path[draw=drawColor,line width= 0.4pt,line join=round,line cap=round] (138.47, 61.20) -- (138.47, 55.20);

\path[draw=drawColor,line width= 0.4pt,line join=round,line cap=round] (193.58, 61.20) -- (193.58, 55.20);

\node[text=drawColor,anchor=base,inner sep=0pt, outer sep=0pt, scale=  1.00] at ( 83.37, 39.60) {T1};

\node[text=drawColor,anchor=base,inner sep=0pt, outer sep=0pt, scale=  1.00] at (138.47, 39.60) {T2};

\node[text=drawColor,anchor=base,inner sep=0pt, outer sep=0pt, scale=  1.00] at (193.58, 39.60) {Control};

\path[draw=drawColor,line width= 0.4pt,line join=round,line cap=round] ( 49.20, 66.48) -- ( 49.20,198.47);

\path[draw=drawColor,line width= 0.4pt,line join=round,line cap=round] ( 49.20, 66.48) -- ( 43.20, 66.48);

\path[draw=drawColor,line width= 0.4pt,line join=round,line cap=round] ( 49.20, 99.48) -- ( 43.20, 99.48);

\path[draw=drawColor,line width= 0.4pt,line join=round,line cap=round] ( 49.20,132.47) -- ( 43.20,132.47);

\path[draw=drawColor,line width= 0.4pt,line join=round,line cap=round] ( 49.20,165.47) -- ( 43.20,165.47);

\path[draw=drawColor,line width= 0.4pt,line join=round,line cap=round] ( 49.20,198.47) -- ( 43.20,198.47);

\node[text=drawColor,anchor=base,inner sep=0pt, outer sep=0pt, scale=  1.00] at ( 34.80, 64.48) {1};

\node[text=drawColor,anchor=base,inner sep=0pt, outer sep=0pt, scale=  1.00] at ( 34.80, 95.48) {2};

\node[text=drawColor,anchor=base,inner sep=0pt, outer sep=0pt, scale=  1.00] at ( 34.80,128.47) {3};

\node[text=drawColor,anchor=base,inner sep=0pt, outer sep=0pt, scale=  1.00] at ( 34.80,161.47) {4};

\node[text=drawColor,anchor=base,inner sep=0pt, outer sep=0pt, scale=  1.00] at ( 34.80,193.47) {5};
\end{scope}
\begin{scope}
\path[clip] (  0.00,  0.00) rectangle (252.94,252.94);
\definecolor{drawColor}{RGB}{0,0,0}

\node[text=drawColor,anchor=base,inner sep=0pt, outer sep=0pt, scale=  1.00] at (138.47, 15.60) {Group};

\node[text=drawColor,rotate= 90.00,anchor=base,inner sep=0pt, outer sep=0pt, scale=  1.00] at ( 10.80,132.47) {Perceived deception};
\end{scope}
\begin{scope}
\path[clip] (  0.00,  0.00) rectangle (252.94,252.94);
\definecolor{drawColor}{RGB}{0,0,0}

\path[draw=drawColor,line width= 0.4pt,line join=round,line cap=round] ( 49.20, 61.20) --
	(227.75, 61.20) --
	(227.75,203.75) --
	( 49.20,203.75) --
	( 49.20, 61.20);
\end{scope}
\begin{scope}
\path[clip] ( 49.20, 61.20) rectangle (227.75,203.75);
\definecolor{fillColor}{RGB}{1,0,0}

\path[fill=fillColor] ( 81.12,154.91) --
	( 83.37,157.16) --
	( 85.62,154.91) --
	( 83.37,152.66) --
	cycle;

\path[fill=fillColor] (136.22,156.99) --
	(138.47,159.24) --
	(140.72,156.99) --
	(138.47,154.74) --
	cycle;

\path[fill=fillColor] (191.33,137.34) --
	(193.58,139.59) --
	(195.83,137.34) --
	(193.58,135.09) --
	cycle;
\end{scope}
\end{tikzpicture}
		\vspace{-8ex}
	\end{center}
	\caption{\label{fig:boxplot_deception} Perceived deception (PDE) by treatment. The $t$-test shows that PDE is significantly higher in T1 than in the control group ($t(96.82)=2.24, p < 0.05, \operatorname{~Cohen's}~d = 0.44 $).}   
\end{figure}
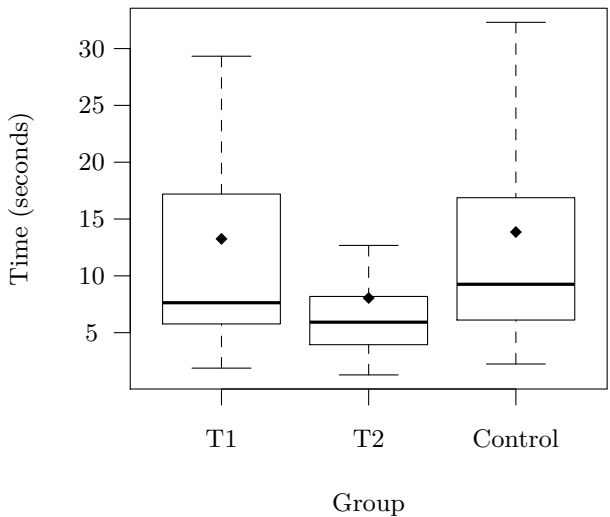
\begin{figure}[t!]
	\begin{center}
		\vspace{-10ex}
\begin{tikzpicture}[x=1pt,y=1pt]
\definecolor{fillColor}{RGB}{1,1,1}
\path[use as bounding box,fill=fillColor,fill opacity=0.00] (0,0) rectangle (252.94,252.94);
\begin{scope}
\path[clip] ( 49.20, 61.20) rectangle (227.75,203.75);
\definecolor{drawColor}{RGB}{0,0,0}

\path[draw=drawColor,line width= 1.2pt,line join=round] ( 61.32, 93.50) -- (105.41, 93.50);

\path[draw=drawColor,line width= 0.4pt,dash pattern=on 4pt off 4pt ,line join=round,line cap=round] ( 83.37, 69.02) -- ( 83.37, 85.56);

\path[draw=drawColor,line width= 0.4pt,dash pattern=on 4pt off 4pt ,line join=round,line cap=round] ( 83.37,185.79) -- ( 83.37,134.19);

\path[draw=drawColor,line width= 0.4pt,line join=round,line cap=round] ( 72.34, 69.02) -- ( 94.39, 69.02);

\path[draw=drawColor,line width= 0.4pt,line join=round,line cap=round] ( 72.34,185.79) -- ( 94.39,185.79);

\path[draw=drawColor,line width= 0.4pt,line join=round,line cap=round] ( 61.32, 85.56) --
	(105.41, 85.56) --
	(105.41,134.19) --
	( 61.32,134.19) --
	( 61.32, 85.56);

\path[draw=drawColor,line width= 1.2pt,line join=round] (116.43, 86.22) -- (160.52, 86.22);

\path[draw=drawColor,line width= 0.4pt,dash pattern=on 4pt off 4pt ,line join=round,line cap=round] (138.47, 66.48) -- (138.47, 77.79);

\path[draw=drawColor,line width= 0.4pt,dash pattern=on 4pt off 4pt ,line join=round,line cap=round] (138.47,114.96) -- (138.47, 95.85);

\path[draw=drawColor,line width= 0.4pt,line join=round,line cap=round] (127.45, 66.48) -- (149.49, 66.48);

\path[draw=drawColor,line width= 0.4pt,line join=round,line cap=round] (127.45,114.96) -- (149.49,114.96);

\path[draw=drawColor,line width= 0.4pt,line join=round,line cap=round] (116.43, 77.79) --
	(160.52, 77.79) --
	(160.52, 95.85) --
	(116.43, 95.85) --
	(116.43, 77.79);

\path[draw=drawColor,line width= 1.2pt,line join=round] (171.54,100.41) -- (215.62,100.41);

\path[draw=drawColor,line width= 0.4pt,dash pattern=on 4pt off 4pt ,line join=round,line cap=round] (193.58, 70.57) -- (193.58, 87.02);

\path[draw=drawColor,line width= 0.4pt,dash pattern=on 4pt off 4pt ,line join=round,line cap=round] (193.58,198.47) -- (193.58,132.82);

\path[draw=drawColor,line width= 0.4pt,line join=round,line cap=round] (182.56, 70.57) -- (204.60, 70.57);

\path[draw=drawColor,line width= 0.4pt,line join=round,line cap=round] (182.56,198.47) -- (204.60,198.47);

\path[draw=drawColor,line width= 0.4pt,line join=round,line cap=round] (171.54, 87.02) --
	(215.62, 87.02) --
	(215.62,132.82) --
	(171.54,132.82) --
	(171.54, 87.02);
\end{scope}
\begin{scope}
\path[clip] (  0.00,  0.00) rectangle (252.94,252.94);
\definecolor{drawColor}{RGB}{0,0,0}

\path[draw=drawColor,line width= 0.4pt,line join=round,line cap=round] ( 83.37, 61.20) -- (193.58, 61.20);

\path[draw=drawColor,line width= 0.4pt,line join=round,line cap=round] ( 83.37, 61.20) -- ( 83.37, 55.20);

\path[draw=drawColor,line width= 0.4pt,line join=round,line cap=round] (138.47, 61.20) -- (138.47, 55.20);

\path[draw=drawColor,line width= 0.4pt,line join=round,line cap=round] (193.58, 61.20) -- (193.58, 55.20);

\node[text=drawColor,anchor=base,inner sep=0pt, outer sep=0pt, scale=  1.00] at ( 83.37, 39.60) {T1};

\node[text=drawColor,anchor=base,inner sep=0pt, outer sep=0pt, scale=  1.00] at (138.47, 39.60) {T2};

\node[text=drawColor,anchor=base,inner sep=0pt, outer sep=0pt, scale=  1.00] at (193.58, 39.60) {Control};

\path[draw=drawColor,line width= 0.4pt,line join=round,line cap=round] ( 49.20, 82.29) -- ( 49.20,188.65);

\path[draw=drawColor,line width= 0.4pt,line join=round,line cap=round] ( 49.20, 82.29) -- ( 43.20, 82.29);

\path[draw=drawColor,line width= 0.4pt,line join=round,line cap=round] ( 49.20,103.56) -- ( 43.20,103.56);

\path[draw=drawColor,line width= 0.4pt,line join=round,line cap=round] ( 49.20,124.84) -- ( 43.20,124.84);

\path[draw=drawColor,line width= 0.4pt,line join=round,line cap=round] ( 49.20,146.11) -- ( 43.20,146.11);

\path[draw=drawColor,line width= 0.4pt,line join=round,line cap=round] ( 49.20,167.38) -- ( 43.20,167.38);

\path[draw=drawColor,line width= 0.4pt,line join=round,line cap=round] ( 49.20,188.65) -- ( 43.20,188.65);

\node[text=drawColor,anchor=base,inner sep=0pt, outer sep=0pt, scale=  1.00] at ( 34.80, 79.29) {5};

\node[text=drawColor,anchor=base,inner sep=0pt, outer sep=0pt, scale=  1.00] at ( 34.80,100.56) {10};

\node[text=drawColor,anchor=base,inner sep=0pt, outer sep=0pt, scale=  1.00] at ( 34.80,121.84) {15};

\node[text=drawColor,anchor=base,inner sep=0pt, outer sep=0pt, scale=  1.00] at ( 34.80,143.11) {20};

\node[text=drawColor,anchor=base,inner sep=0pt, outer sep=0pt, scale=  1.00] at ( 34.80,164.38) {25};

\node[text=drawColor,anchor=base,inner sep=0pt, outer sep=0pt, scale=  1.00] at ( 34.80,185.65) {30};
\end{scope}
\begin{scope}
\path[clip] (  0.00,  0.00) rectangle (252.94,252.94);
\definecolor{drawColor}{RGB}{0,0,0}

\node[text=drawColor,anchor=base,inner sep=0pt, outer sep=0pt, scale=  1.00] at (138.47, 15.60) {Group};

\node[text=drawColor,rotate= 90.00,anchor=base,inner sep=0pt, outer sep=0pt, scale=  1.00] at ( 10.80,132.47) {Time (seconds)};
\end{scope}
\begin{scope}
\path[clip] (  0.00,  0.00) rectangle (252.94,252.94);
\definecolor{drawColor}{RGB}{0,0,0}

\path[draw=drawColor,line width= 0.4pt,line join=round,line cap=round] ( 49.20, 61.20) --
	(227.75, 61.20) --
	(227.75,203.75) --
	( 49.20,203.75) --
	( 49.20, 61.20);
\end{scope}
\begin{scope}
\path[clip] ( 49.20, 61.20) rectangle (227.75,203.75);
\definecolor{fillColor}{RGB}{1,0,0}

\path[fill=fillColor] ( 81.12,117.40) --
	( 83.37,119.65) --
	( 85.62,117.40) --
	( 83.37,115.15) --
	cycle;

\path[fill=fillColor] (136.22, 95.27) --
	(138.47, 97.52) --
	(140.72, 95.27) --
	(138.47, 93.02) --
	cycle;

\path[fill=fillColor] (191.33,119.96) --
	(193.58,122.21) --
	(195.83,119.96) --
	(193.58,117.71) --
	cycle;
\end{scope}
\end{tikzpicture}
		\vspace{-8ex}
	\end{center}
	\caption{\label{fig:boxplot_time} Time spent on responding to the consent dialog by treatment. The KW-test shows a significant difference between T1 and T2 ($\chi^2(1)=8.89$, $p<0.005$).}
\end{figure}

\subsection{Post-hoc Analyses}
\label{ssec:posthoc}

At the beginning of the questionnaire, we asked participants to recall which purposes they have agreed to in the consent dialog. Thus, we are able to compare the accuracy of participants' statements between groups. As reported in Table~\ref{tab:hypotheses_tests}, the difference is highly significant between all three groups ($\chi^2(2)=11.01$, $p<0.005$). 

When looking at the proportion of participants who declined all purposes, it is notable that 50\% of the control group, but only 32\% of the deception group chose this option. 
After informing participants about their choice, we specifically asked those who agreed to at least one purpose whether they had been aware of the possibility to decline all purposes.  Only 32\% stated to be aware of this option. However, the proportion of aware participants does not differ significantly between the deception and control group ($\chi^2(1)=2.71$, $p=0.10$)\unskip. It seems that even the design of our control dialog, possibly in combination with learned expectations, imposes some pressure to select at least one option on a subset of the participants. This highlights that future research could seek to improve the communication of the ``freely given'' aspect of GDPR-compliant consent (cf.~Section~\ref{sec:legal}).

To test whether users' privacy attitudes regarding cookies influence their reaction to the cookie dialog, we also test the relationship between the number of chosen purposes and PA. For this analysis we only consider the groups that were presented all three purposes. We find a weak but significant \emph{negative} correlation between privacy attitudes and the number of consented purposes ($r_{s}=-0.23$, $p<0.05$, $n=102$). However, the difference in PA between those who clicked the deceptive button and those who did not, is not significant ($t(91.9)=-1.04$, $p=0.30$, $n=98$). Moreover, as expected, privacy attitudes do not differ significantly between groups as participants were randomly assigned to groups. This reassures us that the PA items measure trait rather than state.

\section{Discussion}
\label{sec:discussion}
Next we reflect on the results, 
then discuss limitations (Section~\ref{ssec:robustness}), and comment on recent developments in the space (Section~\ref{ssec:recent}).

\subsection{Summary and Interpretation}
\label{ssec:summary}

Our experimental results confirm the common conjecture that design elements of consent dialogs can nudge users towards making specific choices. We show empirically that the selection of data processing purposes, as required by the GDPR, is not exempt: users accept more data collection purposes when consent dialogs integrate a highlighted default button that selects all purposes at once. Surprisingly, we observe a four times stronger effect for our multi-purpose consent dialog than previously reported for simple default buttons in binary consent dialogs.
Moreover, the fact that users who click this button are less likely to correctly recall the consented purposes casts doubt on the morality and legitimacy of this design element, as it might lead users to act against their intention. This interpretation is further supported by the finding that users tend to regret their decision after being informed about the effective purposes.

Besides the effect of deceptive default buttons, we present more encouraging results on the possibility of differentiating between consent decision for multiple purposes in one dialog: although the number of purposes significantly affects the response time, the difference in perceived difficulty is insignificant. This indicates that most users can handle three different purposes without experiencing the negative effects predicted by the theory of choice proliferation. Of course, more research is needed to investigate  the critical number of purposes. Also choice structure, the other relevant determinant in choice proliferation, requires further attention~\cite{korff2014too}.

Our analysis of control variables reveals that users with stronger stated privacy attitudes consent to fewer purposes. While this result seems to challenge the privacy paradox (a term for the often observed discrepancy between stated attitudes and privacy behavior~\cite{norberg2007,sundar2013unlocking,gerber2018}), 
it must be interpreted with caution. First, our instrument is not ideal to study the paradox. It confounds this relationship with the dominant effect of a deceptive default button and measures the privacy attitude only after recalling the effective purposes. Second, unlike in many studies that find the paradox, our items measure privacy attitudes quite narrowly for the specific domain: two out of three items mention cookies. According to the principle of compatibility, behavior is more predictable from attitudes if it is measured on the same level of specificity~\cite{ajzen1998models}. Third, the interpretation of attitude--behavior links is problematic if the behavior is partly unintentional, such as accepting undesired purposes.
To some extent, this corroborates nuanced or critical perspectives on the privacy paradox~\cite{dienlin2015}.

\subsection{Validity and Limitations}
\label{ssec:robustness}

To gauge the relevance of our results, one may ask how prevalent the tested dialog is on the web. Unfortunately, reliable data in this dynamic space is scarce. The most recent data in~\cite{utz2019ccs} refer to a snapshot in August 2018 and thus predate the introduction of the dialog on the airline website, where we discovered it, and possibly elsewhere. According to this snapshot, only 7\% of web consent notices are blocking, and 8\% present multiple purposes (immediately or on request). These shares almost certainly increased with the adoption of consent managers in the course of 2019 (see Sect.~\ref{ssec:recent} below). 

\begin{table}
	\centering
	\caption{Robustness of the main effects: $p$-values of hypothesis tests broken down by the location of the classroom experiment.}
	\label{tab:dif_DE_AT}
	\begin{tabular}{lrr}
	\textbf{Hypothesis and contrast groups}
		 & \textbf{Austria} & \textbf{Germany} \\ 
		 & \small ($n=90$) & \small ($n=60$)  \\
		 \midrule
	 H1: T1 vs Control                 & 0.028          & 0.090         \\
	 H2a: RE-before/-after in T1 & 0.036          & 0.098         \\
	 H2b: T1 vs Control                & 0.019          & 0.564         \\
	 H3: T1 vs T2                      & 0.031          & 0.041         \\
	 H4: T1 vs T2  (rejected)                     & 0.572          & 0.560         \\
	\end{tabular}\\[1ex]
\end{table}

However, our choice of stimulus was not driven by the most prevalent design, which we and other researchers~\cite{trevisan2019pets,Sanchez-Rola2019,utz2019ccs} suspect to trivially violate the GDPR. Instead, we set out to study elementary design options of multi-purpose dialogs, the novel and most under-researched aspect of consent dialogs. Our dialog implements opt-in and does not proceed without an affirmative action (blocking) in order to anticipate future good practices. The fact that similar dialogs are used by respectable organizations with competent legal departments and millions of unique users per year\footnote{Figures extracted from media data of the airline's online services, available from the authors on request.} adds to the relevance. More importantly, since we study individual effects derived from theory, the prevalence of our stimulus material is of subordinate importance. We aim to identify generalizable effects, which could be studied on real or artificial dialogs. The choice of using a real dialog for inspiration along with a credible cover story is merely one of multiple measures to assure the external validity of our lab study. 

To check for possible risks to external validity, we analyzed the participants' free-text responses for prejudiced assumptions about our study. Only one participant exhibited demand characteristics~\cite{orne1962social}.
The person wrote that he or she has agreed to all purposes because the website was part of a scientific experiment. All remaining participants answered as if they were dealing with an actual flight search website. Moreover, we do not find further indications that the participants might have perceived our stimuli as artificial. 

It is important to mention that the study has limitations. First, the experimental setup may not fully reflect users' actual behavior regarding consent dialogs. Even though we made an effort to hide the research purpose of our study, we cannot rule out that participants might have guessed our focus on cookie choices or privacy in general. Moreover, our sample is limited to German-speaking computer science students who are probably more educated about the functionality of cookies and the web in general. This \emph{known} bias, however, does not compromise the upshot of this paper: if even computer literate populations fall for the deceptive design, we must assume that the outcome for the general public is even worse. We tried to mitigate all other disadvantages of convenience samples by replicating the experiment in two geographically distant universities.
Table~\ref{tab:dif_DE_AT} confirms that H1--H3 are supported in both populations.\footnote{Some $p$-values for Germany are above 5\% only because we conservatively use the two-sided test. The values for the one-sided test are half of the reported ones. Recall that group sizes in Germany alone may be below 20 subjects.}
We chose a classroom experiment (and accepted  its limitations) in order to reduce \emph{unknown} biases due to participant self-selection, which is an acute problem of empirical privacy research~\cite{walsh1992poq,cho1999privacy}. Precisely the attitudes and beliefs of interest correlate with non-response and dropouts. For perspective, our completion rate is above 88\% in all sessions, whereas the concurrent field experiment received 110 completed surveys from more than 30,000 solicitations, translating to a response rate below 0.4\%~\cite{utz2019ccs}. (The authors acknowledge this bias and  chose not to analyze self-reported data quantitatively.)

\begin{figure}[t]
	\begin{center}
	\includegraphics[width=5.5cm]{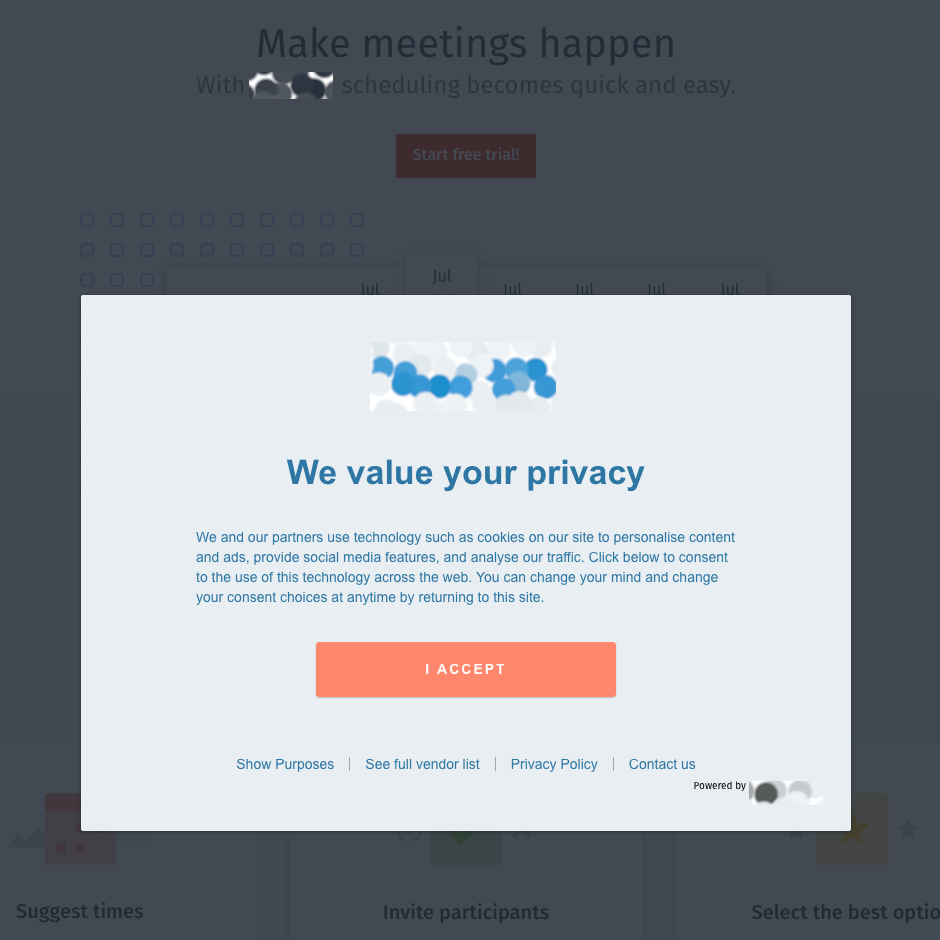}
	\caption{\label{fig:doodle} Dialog of a commercial consent manager (August 2019).}
	\end{center}
\end{figure}
\subsection{Recent Developments}
\label{ssec:recent}

A simple interface adjustment, meanwhile implemented in the airline website, is to change the button text from ``Select all and confirm'' to ``Select and confirm'' (and change the function accordingly) as soon as the first checkbox is selected. While this breaks with the design principle that button semantics should be stateless, it might avoid the severest mishaps where cognitive effort that went into selecting purposes is wasted. It requires another user study with more participants to gauge if this modification reduces the disappointment in the subset of users who select at least one but not all purposes.\footnote{This description applied to the time of writing in mid-2019. The checkbox logic had been changed once again when we revisited the website in fall 2019 for the preparation of the camera-ready version. This highlights the dynamics in this space.}

In the past months, we (and others \cite{Sanchez-Rola2019}) have observed other ``innovative'' consent dialogs, such as page-long lists of affiliate partners for third-party tracking, that call for tailored user studies through the lenses of deception and choice proliferation. For example, a popular meeting scheduling service uses a modal dialog entitled ``We value your privacy'' with a prominent button labeled ``I accept.'' To access literally hundreds of options, one has to click on ``Show purposes'', which is a text link next to three others (see Fig.~\ref{fig:doodle}). Interestingly, this dialog seems to be operated (and presumably evaluated) by an intermediary specialized on consent management. This fits into the picture where ENISA, a EU agency, mentions ``consent management'' as a new business opportunity for cybersecurity startups~\cite[p.~10]{ENISA2019}.

\section{Conclusion}
\label{sec:conclusion}

This study presents new empirical evidence supporting that design elements used in consent dialogs of popular websites might deceive users into agreeing to more data processing purposes than intended. It complements the measurement studies~\cite{degeling2018we,trevisan2019pets,vanEijk2019,Sanchez-Rola2019} that emphasize the wide adoption of such ``dark patterns''~\cite{bosch2016tales} as well as a recent field study on cookie banners~\cite{utz2019ccs}. Based on these findings, we can derive recommendations for user interface designers and policy makers.

Our first recommendation reiterates calls respect the user's interest: instead of nudging users towards agreements that mainly benefit the party who owns the website, defaults should reflect either a privacy-aware safe choice or elicit the majority's preferences as a descriptive social norm. This could be achieved by designing a set of best-practice consent dialogs, incorporating the body of knowledge from behavioral privacy research. These templates can be made available to organizations who value consumers' privacy or seek legal certainty without commissioning an intermediary.

However, past and ongoing efforts in the usable privacy research community towards understanding how to nudge users into making safer choices are void if the industry tries to achieve the opposite. Since the value of personal data increases with the number of possible secondary uses~\cite{SABH2015}, businesses have incentives to maximize the number of consented purposes. It is tempting to call for a regulator or oversight body to step in and ensure that dialogs are designed in the users' interest. But we are hesitant about suggesting more (or more specific) regulations for two reasons. First, the GDPR stipulates freely given, unambiguous, and informed consent. It may take a court decision to provide clarity over the fact that the practices we observe \emph{do not} meet these requirements and hence \emph{cannot} provide a legal basis for personal data processing. However, such decisions must be based on further empirical research. Second, the time and cognitive effort millions of users regularly spend on consent dialogs may not justify the outcome at the societal level. Rather than mandating special forms of consent dialogs (which hardly work for devices without display or network services that are not customer-facing), a policy priority should be the establishment of a standard for non-interactive privacy preference negotiations. 

It seems that P3P~\cite{cranor2003p3p} was 20 years ahead of its time, and the do-not-track header too simple and polarized~\cite{Olejni2019}. There could be a middle ground in which consent dialogs do not disappear. But their design moves from the hands of data controllers to developers of user agents, who compete for the best service in the data subject's interest. In order to foster competition, and not to repeat the mistakes of do-not-track, it is important that browser and app vendors must be required to interoperate with any privacy agent of the user's choice.

\section*{Acknowledgements}
We thank all participants of the pretests and the main study. We also thank Daniel Woods, Henry Hosseini and our shepherd Blase Ur for useful discussions as well as the anonymous reviewers for many constructive comments. This work received funding from the German Bundesministerium für Bildung und Forschung (BMBF) under grant agreement 16KIS0382 (AppPETs) and the Archimedes Privatstiftung, Innsbruck.

\section{Appendix}
\label{sec:appendix}

\begin{figure}[h!]
	\begin{center}
		\includegraphics[width=5.5cm]{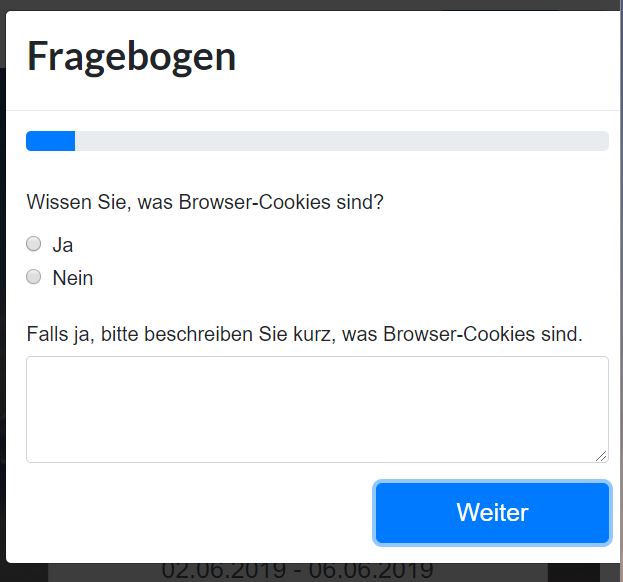}
	\end{center}
	\caption{\label{fig:questionnaire_german} Pop-up with questionnaire.}
\end{figure}

\begin{figure*}[tbp]
	\begin{center}
		\begin{tabular}{ccc}
			\includegraphics[width=5.5cm]{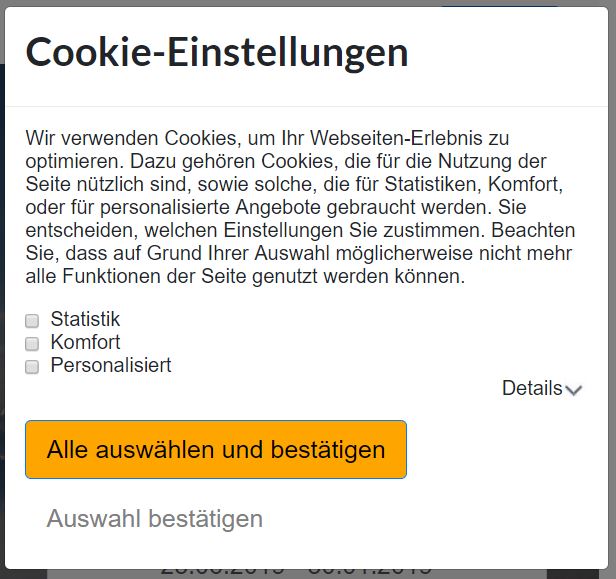} & 
			\includegraphics[width=5.5cm]{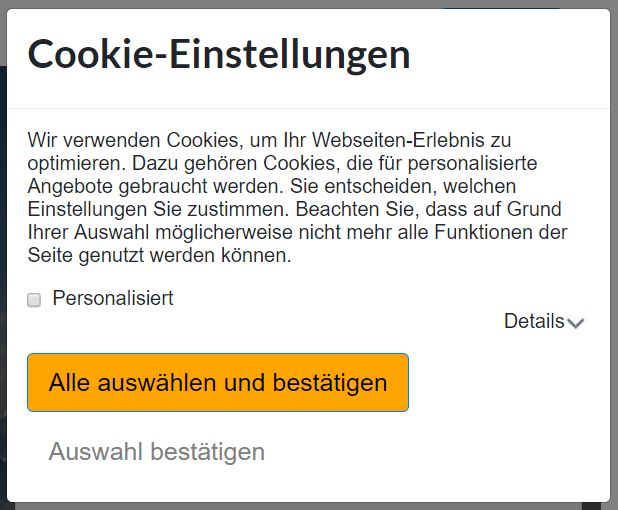} & 
			\includegraphics[width=5.5cm]{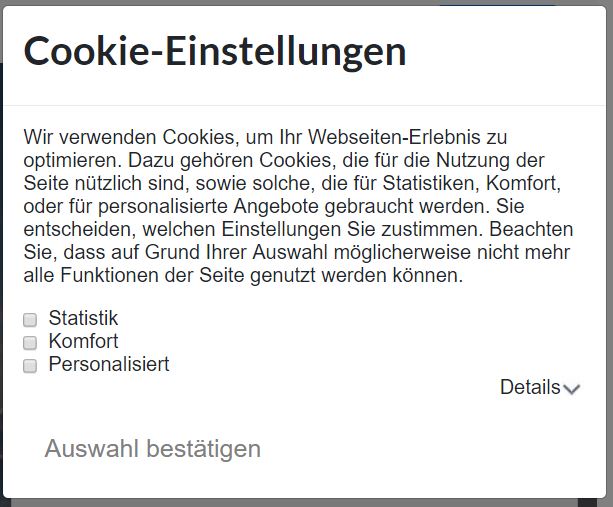} \\
			Deception (T1)  & Reduced choice (T2)   & Control \\                             
		\end{tabular}
	\end{center}
	\caption{\label{fig:banner_versions_german} Original German version of the stimulus material.}
\end{figure*}

\begin{figure}[tb]
	\begin{center}
		\includegraphics[width=4cm]{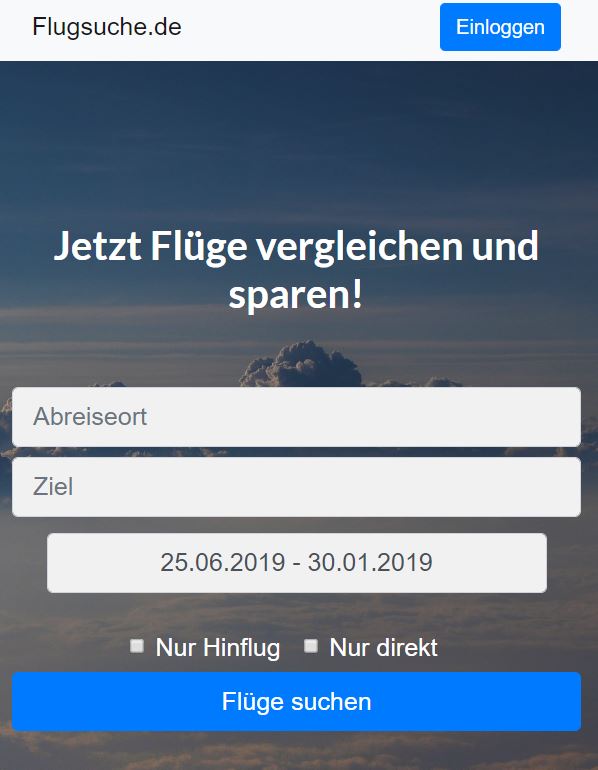}
	\end{center}
	\caption{\label{fig:website_german} Functional mock-up website offering flight search.}
\end{figure}

\begin{figure}[h!]
	\begin{center}
	\includegraphics[width=5.5cm]{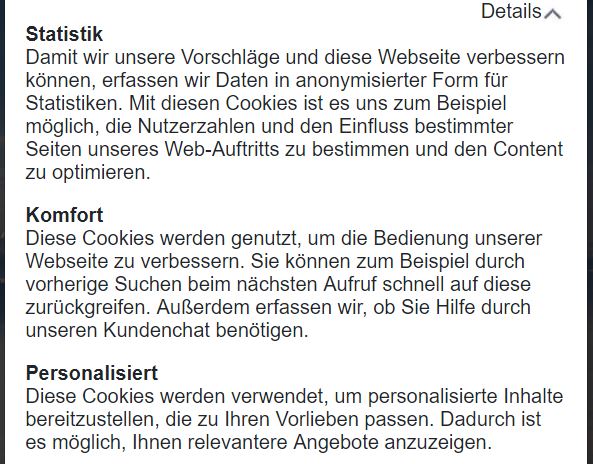}
	\end{center}
	\caption{\label{fig:details_german} German version of the expanded cookie dialog after clicking on ``details''.}
\end{figure}

\begin{table}[htb]
	\centering
	\caption{Constructs and corresponding items in German. }
	\label{tab:construct_items_german}
	\begin{tabular}{p{1.4cm}p{6.3cm}}
		{\textbf{Item}} & {\textbf{Item text (original)}} \\ \midrule
		\multicolumn{2}{@{}p{7.5cm}}{\textit{Perceived Deception (PDE)}}  \\
		\hspace{0.18cm}PDE1 & \GERPDEa   \\ 
		\hspace{0.18cm}PDE2 & \GERPDEb   \\
		\hspace{0.18cm}PDE3 & \GERPDEc   \\[1ex]
		\multicolumn{2}{@{}p{7.5cm}}{\textit{Perceived Difficulty (PDI)}}\\
		\hspace{0.18cm}PDI1 & \GERPDIa \\ 
		\hspace{0.18cm}PDI2 & \GERPDIb \\ 
		\hspace{0.18cm}PDI3$\leftrightarrow$ & \GERPDIc \\[1ex]
		\multicolumn{2}{@{}p{7.5cm}}{\textit{Regret (RE)}}  \\ 
		\hspace{0.18cm}RE1 & \GERRGa \\ 
		\hspace{0.18cm}RE2 & \GERRGb \\ 
		\hspace{0.18cm}RE3$\leftrightarrow$ & \GERRGc \\[1ex] 	
		\multicolumn{2}{@{}p{7.5cm}}{\textit{Privacy Attitudes (PA)}}  \\ 
		\hspace{0.18cm}PA1 & \GERPAa \\ 
		\hspace{0.18cm}PA2 & \GERPAb \\ 
		\hspace{0.18cm}PA3 & \GERPAc \\
	\end{tabular}\\[1ex]
	\small
	Items marked with `$\leftrightarrow$' use inverted scales.
\end{table}

\end{document}